\documentclass[11pt,twocolumn,aps,prl,superscriptaddress,reprint,citeautoscript]{revtex4-1}

\setcitestyle{super}
\usepackage[abs]{overpic}
\usepackage{graphicx,graphics}
\usepackage{dcolumn}
\usepackage{latexsym,verbatim}
\usepackage{bm}
\usepackage{color}
\usepackage{ulem}
\usepackage{sidecap}
\usepackage[framemethod=default]{mdframed}
\usepackage[breaklinks=true,colorlinks=false,citecolor=blue,linkcolor=blue,urlcolor=blue]{hyperref}
\usepackage[makeroom]{cancel}
\usepackage{fancybox}
\usepackage{physics}

\newcommand{\DAB}[1]{\textcolor{black}{#1}}
\newcommand{\addIP}[1]{\textcolor{BurntOrange}{#1}}

\newcommand{\blue}[1]{\textcolor{black}{#1}}

\usepackage{gensymb}
\usepackage{wrapfig}

\usepackage[dvipsnames]{xcolor}
\definecolor{SIcolor}{RGB}{219, 48, 122}

\definecolor{DAB}{RGB}{255, 255, 0}

\begin{document}
\author{I.Y. Phinney}
\affiliation{Massachusetts Institute of Technology, Cambridge, MA 02139}
\author{D.A. Bandurin$^{*}$}
\affiliation{Massachusetts Institute of Technology, Cambridge, MA 02139}
\author{C. Collignon}
\affiliation{Massachusetts Institute of Technology, Cambridge, MA 02139}
\author{I.A. Dmitriev}
\affiliation{Physics Department, University of Regensburg, 93040, Regensburg, Germany}
\affiliation{Ioffe Institute, 194021 St. Petersburg, Russia}
\author{T.~Taniguchi}
\affiliation{International Center for Materials Nanoarchitectonics, National Institute of Material Science, Tsukuba 305-0044, Japan}
\author{K.~Watanabe}
\affiliation{Research Center for Functional Materials, National Institute of Material Science, Tsukuba 305-0044, Japan}
 \author{P. Jarillo-Herrero$^{*}$}
\affiliation{Massachusetts Institute of Technology, Cambridge, MA 02139}

\title{\blue{Strong interminivalley scattering in twisted bilayer graphene revealed \\ by high-temperature magnetooscillations}}

\begin{abstract}
Twisted bilayer graphene (TBG) provides an example of a system in which the interplay of interlayer interactions and superlattice structure impacts electron transport in a variety of non-trivial ways and  gives rise to a plethora of interesting effects. Understanding the mechanisms of electron scattering in TBG has, however, proven challenging, raising many questions about the origins of resistivity in this system. Here we show that TBG exhibits high-temperature magnetooscillations originating from the scattering of charge carriers between TBG minivalleys. The amplitude of these oscillations reveals that interminivalley scattering is strong, and its characteristic time scale is comparable to that of its intraminivalley counterpart. Furthermore, by exploring the temperature dependence of these oscillations, we estimate the electron-electron collision rate in TBG and find that it exceeds that of monolayer graphene. Our study demonstrates the consequences of the relatively small size of the superlattice Brillouin zone and Fermi velocity reduction on lateral transport in TBG. 
\end{abstract}
\maketitle

\begin{figure*}[ht!]
	\centering
	\includegraphics[width=1\textwidth]{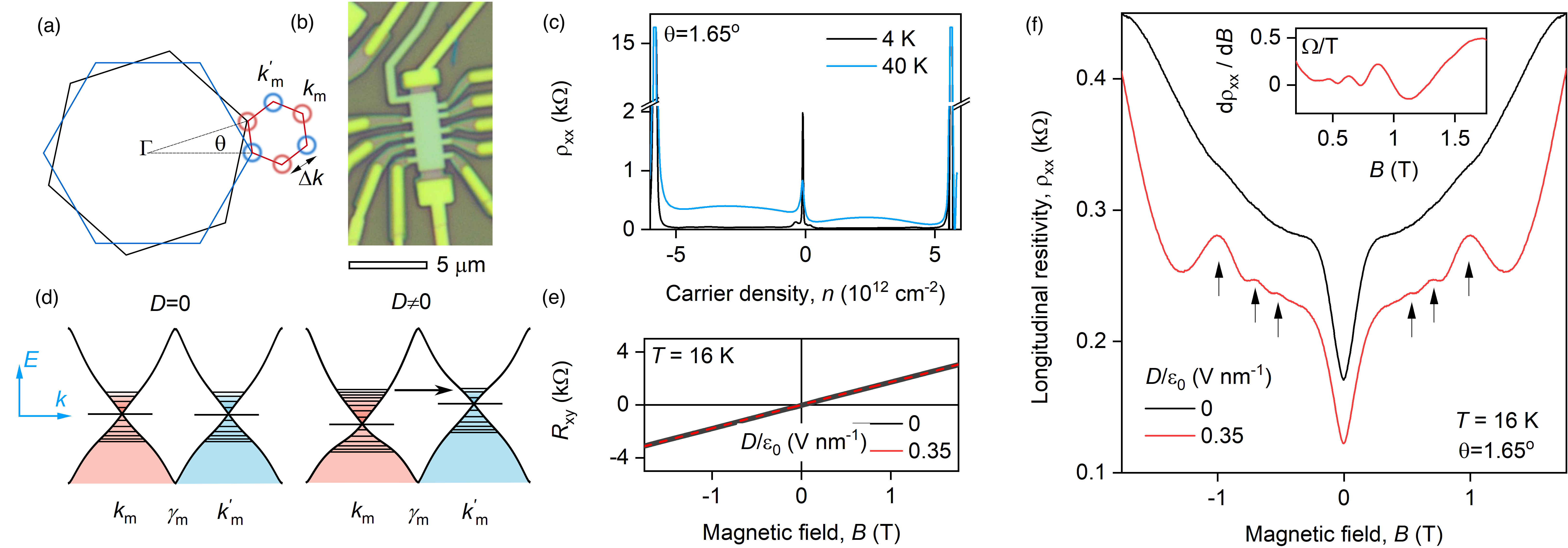}
	\caption{Interminivalley magnetooscillations in small-angle twisted bilayer graphene. (a) Schematic illustration of the mini-BZ of the SA-TBG superlattice. Red and blue circles represent Fermi surfaces in different minivalleys, labelled $k_\mathrm{m}$ and $k_\mathrm{m}'$.  (b) Optical photograph of an encapsulated SA-TBG device. Yellow - gold contacts, dull green - top gate, light brown - Hall bar mesa. (c) $\rho_\mathrm{xx}$ as a function of $n$ for the $1.65\degree$ device at $4$~K (black) and $40$~K (blue). $B=0$, $D=0$. (d) Calculated single-particle band structure for the 1.65$\degree$ SA-TBG: at low-energies two Dirac cones are formed in the vicinity of the $k_\mathrm{m}$ and $k'_\mathrm{m}$ points. The horizontal black lines represent unevenly spaced LLs that form in the presence of a perpendicular magnetic field. When $D\neq 0$, the cones are shifted with respect to each other. When LLs from different minivalleys get aligned inside the thermal window around the Fermi level, interminivalley scattering (arrow) is enhanced resulting in excess resistivity.  (e) Hall resistance, $R_\mathrm{xy}$, as a function of $B$ for two characteristic $D=0$ and $D/\varepsilon_0=0.35$~V/nm yielding $n\approx0.7\times10^{12}$~cm$^{-2}$ measured at $T=16$~K.  (f) $\rho _\mathrm{xx}$ (symmetrized) as a function of $B$ for the same $D$, $n$ and $T$ as in (e). Inset: Derivative of the $\rho_\mathrm{xx}(B)$ dependence for the case of $D/\varepsilon_0=0.35~$V/nm.}
	\label{fig:Fig1}
\end{figure*}

Two graphene monolayers, placed on top of each other with a small rotational misalignment between their crystallographic axes, form a long-wavelength moiré superlattice. The electronic properties of such a superlattice depend on the relative twist angle, $\theta$, between the graphene layers as well as their interlayer hybridization. A particularly interesting case is that of small-angle ($\theta<3\degree$) TBG (SA-TBG), where hybridization is strong, and which, for a certain range of angles, features intriguing interaction-driven phenomena including, but not limited to, superconductivity~\cite{Yuan_SC,Yankowitz}, correlated insulator states~\cite{Yuan_Insulators}, and orbital ferromagnetism~\cite{sharpe2019, serlin2020}. The low-energy single-particle band structure of SA-TBG resembles that of monolayer graphene (MLG) but is characterized by a decreased Fermi velocity, $v_\mathrm{F}$, and a reduced Brillouin zone (BZ)~\cite{SantosTBG}. Like the BZ of MLG, the reduced BZ is hexagonal and contains two minivalleys located at the $k_\mathrm{m}$ and $k'_\mathrm{m}$ high-symmetry points~\cite{bistritzer2011}. The minivalleys are spaced apart by a relatively small (in comparison to MLG) distance\addIP{,} $\Delta k=(4\pi/a)\sin(\theta/2)$, where $a$ is the lattice constant of MLG (Fig.~\ref{fig:Fig1}a). In MLG, the intervalley separation is sufficiently large so as to suppress   intervalley electron scattering, provided that atomically-sharp defects are absent~\cite{CastroNetoRMP}. In this work, we show that the opposite is true for SA-TBG, where strong interminivalley scattering significantly affects its transport properties.

Our devices are multi-terminal Hall bars composed of SA-TBG encapsulated between two relatively thin ($<100$~nm thick) slabs of hexagonal boron nitride (hBN). The Hall bars were produced by a combination of tear-and-stack~\cite{Tutuc_TBG,lasercut} and hot release~\cite{Hot-transfer_NComm} methods, were endowed with quasi-one-dimensional contacts~\cite{MosheNatPhys} and had typical width of about $2$ $\mu$m as shown in Fig.~\ref{fig:Fig1}b (See Supplemental Material~\cite{Supplementary}). Figure 1c shows a typical dependence of the longitudinal resistivity, $\rho_\mathrm{xx}$, versus the externally-induced total carrier density $n$, measured in one of our devices at two representative temperatures, $T$. At small $n$, the $\rho_\mathrm{xx}(n)$ dependence resembles that of monolayer graphene (MLG): namely, it exhibits a sharp peak of about 2~k$\Omega$ at the charge neutrality point (CNP) that rapidly drops to $20-50~\Omega$ with increasing $|n|$. Upon further doping,  $\rho_\mathrm{xx}(n)$ exhibits a steep rise at $|n| \approx 6\times 10^{12}$ cm$^{-2}$, which corresponds to full filling of the first superlattice miniband in accord with previous studies on SA-TBG~\cite{SmetTBG,YuanPRL,Tutuc_TBG}. Three devices with $\theta$ of  $1.65\degree, 2.24\degree$ and $2.3\degree$, respectively, were studied --- all exhibiting similar transport characteristics (see Supplemental Material~\cite{Supplementary} for the angle determination procedure). 

A notable feature of SA-TBG is that, by employing a single- or dual-gated device architecture, one can selectively populate the minivalleys by appropriately tuning the top and bottom gate voltages~\cite{Rickhaus_minivalley,berdyugin2020} ($V_\mathrm{tg}$ and $V_\mathrm{bg}$ respectively). This combination defines the relative displacement field between graphene layers, $D=(C_\mathrm{bg}V_\mathrm{bg}-C_\mathrm{tg}V_\mathrm{tg})/2$, and total carrier density, $n=(C_\mathrm{bg}V_\mathrm{bg}+C_\mathrm{tg}V_\mathrm{tg})/e$. Here $C_\mathrm{tg,bg}$ are the top and bottom gate capacitances per unit area,
and $e$ is the electron charge.  Figure~\ref{fig:Fig1}d shows the calculated band structure of the $1.65\degree$ SA-TBG  for the case of zero and finite $D$, which clearly demonstrates the gate-induced imbalance in the population between the $k_\mathrm{m}$ and $k'_\mathrm{m}$ points in the latter case. Intuitively, as the minivalleys are predominantly formed from the energy bands of different graphene sheets, an applied electric field dopes the layers unequally resulting in such an imbalance~\cite{VolodyaScreening,berdyugin2020,Rickhaus_minivalley}.
In the presence of a perpendicular magnetic field, $B$, the gate-induced imbalance also determines the relative offset of the Landau levels (LLs) hosted by each minivalley~\cite{SanchezPRL,FallahazadPRB}, a property that brings us a reliable method to explore the effects of interminivalley electron scattering in SA-TBG, as we now proceed to show.  

Figure~\ref{fig:Fig1}f compares the magnetoresistance of one of our devices measured at $T = 16$~K, for zero and finite   ${D}/{\varepsilon_0}=0.35$~V/nm at the same total $n$ (where $\varepsilon_0$ is the vacuum permittivity). The carrier density was verified via the Hall effect measurements presented in Fig.~\ref{fig:Fig1}e, which shows that the Hall resistance, $R_\mathrm{xy}$, and, therefore, $n$, is identical for both $D$ values. At zero $D$, $\rho_\mathrm{xx}$ grows with increasing $B$ but remains featureless: Shubnikov-de-Haas oscillations (SdHO) in this device disappear at $15~$K for this $B$ range~\cite{YuanPRL} (see below). In striking contrast, a clear oscillatory pattern develops in $\rho_\mathrm{xx}(B)$ data when a finite $D/\varepsilon_0=0.35~$V/nm is applied across the graphene layers. The oscillations are even more visible in the derivative of the resistivity with respect to $B$,  $\mathrm{d}\rho_\mathrm{xx}/\mathrm{d}B$, (inset in Fig.~\ref{fig:Fig1}f) because of the eliminated magnetoresistance background.

\begin{figure*}[ht!]
	\centering
	\includegraphics[width=1\textwidth]{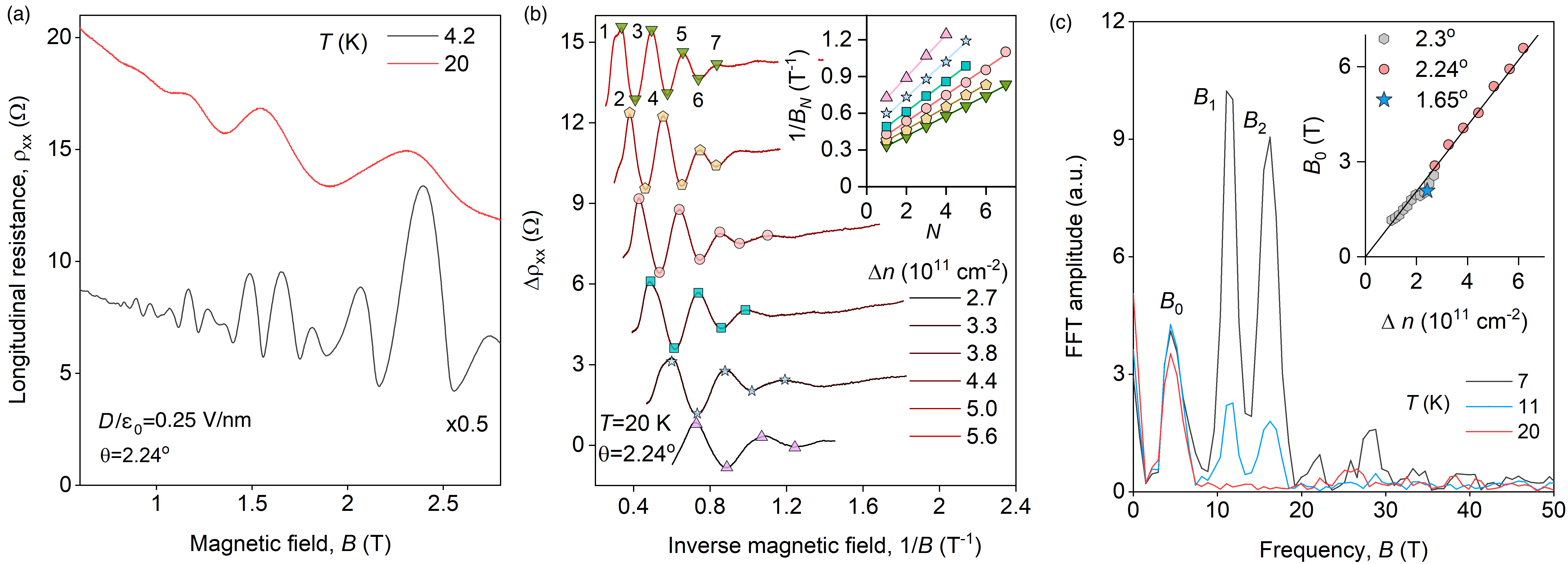}
	\caption{Fundamental frequency of the interminivalley magnetooscillations in SA-TBG. (a) $\rho _\mathrm{xx}$ as a function of $B$ for two characteristic $T=4.2$ and $20$~K measured in $2.24\degree$ SA-TBG. $D/\varepsilon_0=0.25~$V/nm and $n=2.76\times10^{12}$~cm$^{-2}$. 
    (b) Amplitude of interminivalley magnetooscillations $\Delta \rho_\mathrm{xx}$ in our $2.24\degree$ device at $20$~K after the subtraction of a smooth non-oscillating background. Inset: Symbols show resonant values of the inverse magnetic field $1/B_\mathrm{N}$ plotted against $N$ for different $n$ corresponding to the matching symbols in (b). Lines: Fits to the linear dependence yielding $B_\mathrm{0}$ (see text).
	(c) Examples of the FFT spectra of the SA-TBG magnetooscillations for characteristic $T$. $B_\mathrm{1}$ and $B_\mathrm{2}$ determine the SdHO periodicity in different minivalleys and $B_\mathrm{0}$ is the frequency of interminivalley magnetooscillations. Inset: $B_\mathrm{0}$ as a function of population imbalance $\Delta n$ for SA-TBG of different angles. Solid line is the expected $B_\mathrm{0}=h\Delta n/4e$ dependence. }
	\label{fig:Fig2}
\end{figure*}

Figure~\ref{fig:Fig2}a details our observations further by comparing the low-field magnetoresistance of another SA-TBG device ($2.24\degree$) at two characteristic $T$ and $D\neq0$. At $T=4.2$ K, $\rho_\mathrm{xx}$ exhibits the $1/B$-periodic pattern ascribed to SdHO. Because the applied displacement field creates a small difference in the size of the Fermi surfaces associated with the $k_\mathrm{m}$ and $k'_\mathrm{m}$ minivalleys (see Fig~\ref{fig:Fig1}d), two oscillations of slightly different frequency emerge~\cite{VolodyaBeating,SmetTBG,Chung2018}. The sum of these produces a familiar beating pattern.  At $T>20$~K, a different oscillation series, characterized by a much lower frequency, dominates the $\rho_\mathrm{xx}(B)$ behaviour. In Fig.~\ref{fig:Fig2}b we plot the amplitude, $\Delta \rho_\mathrm{xx}$, of these oscillations as a function of the inverse magnetic field $1/B$ and demonstrate 
their $1/B$-periodicity. This periodicity is further verified by plotting the oscillations' extrema indices, $N$, against the values of the inverse magnetic field, $1/B_\mathrm{N}$, at which they appear: all peaks (dips) fall onto straight lines, the slope of which defines the oscillation frequency as $B_\mathrm{0}=\frac{1}{2}[\frac{d (1/B_\mathrm{N})}{d N}]^{-1}$ (inset of Fig.~\ref{fig:Fig2}b).

Having revealed the $1/B-$character of the high-$T$ resistance oscillations, it is instructive to explore these magnetooscillations in SA-TBG by FFT analysis. As expected from the beating pattern, the FFT spectrum at $T=4.2~$K is dominated by two closely-spaced peaks, labeled as $B_\mathrm{1}$ and $B_\mathrm{2}$ (Fig.~\ref{fig:Fig2}c), containing information on the carrier density in each minivalley, $n_\mathrm{1,2}$, via $B_\mathrm{1,2}=n_\mathrm{1,2}h/ge$, where $h$ is Planck's constant and $g=4$ is the minivalley  degeneracy~\cite{SmetTBG}. At $T=20~$K, the FFT spectrum consists of a single peak at $B_\mathrm{0}$ (labeled accordingly) that matches the periodicity determined from the $1/B_\mathrm{N}(N)$ fit (inset of Fig.~\ref{fig:Fig2}b). Interestingly, the $B_\mathrm{0}$ peak is also visible at $T=4.2~$K and $11~$K in the FFT spectra, but, because of the complicated beating pattern in $\rho_\mathrm{xx}(B)$, these oscillations were obscured in previous magnetotransport studies on SA-TBG, whereas in large $\theta>3\degree$ non-encapsulated devices, they were presumably absent~\cite{VolodyaBeating}. Critically, we find that the obtained $B_\mathrm{0}$ is identical to the difference $B_\mathrm{2}-B_\mathrm{1}$  indicating that the period of the high$-T$ magnetooscillations is controlled by the carrier density imbalance, $\Delta n=n_\mathrm{2}-n_\mathrm{1}$, between the minivalleys (see below).

Figure~\ref{fig:Fig3}a shows the $\mathrm{d} \rho_\mathrm{xx}/\mathrm{d} B_\mathrm{0}$ for $\theta=2.24\degree$ mapped onto a $(B,T)$ plane. Such representation allows for a convenient illustration of the evolution of magnetooscillation patterns in SA-TBG as a function of $T$: fast SdHO, clearly visible at liquid helium $T$, vanish at $\sim 15$~K whereas the slow high-$T$ oscillations persist even above $50$~K. We also studied the effect of in-plane dc current, $I_\mathrm{dc}$, on magnetoresistance and found that, in contrast to SdHO, which are readily damped by the application of only $I_\mathrm{dc}\approx10~\mu$A because of Joule heating, the amplitude of the high-$T$ magnetooscillations is resilient to $I_\mathrm{dc}$, up to at least $75~\mu$A. Interestingly, we also observed that upon increasing $I_\mathrm{dc}$, the phase of these oscillations flips several times, additionally distinguishing them from SdHO (see below and Supplemental Material~\cite{Supplementary}).

Taken together, the high-$T$ character, peculiar frequency, and fragile phase identify these oscillations as an SA-TBG analogue of magneto-intersubband oscillations (MISO) discovered in wide quantum wells (QW) and studied in related systems~\cite{Coleridge,Polyanovsky,MISO_old,Raichev2008,  RaichevMISO,kartsovnik2002,Minkov,Vitkalov_Bparallel,RevModPhysVanya}. In QW, the oscillations emerge when a two-dimensional electron system (2DES) occupies two or more energy bands capable of electron exchange~\cite{Coleridge,Polyanovsky,MISO_old,Raichev2008,RevModPhysVanya}. In particular, when the LLs from different subbands become aligned within the thermal window around the Fermi level, elastic interband scattering gives rise to excess resistivity. In the opposite case, when the subbands are misaligned, interband scattering is suppressed. As a result, the resistance experiences $1/B-$periodic oscillations with a period proportional to the difference in filling factors between the two subbands.
In the assumption that the intraband scattering time, $\tau$, does not depend on the subband index,  the oscillations' functional form in the limit of small $I_\mathrm{dc}$ reads~\cite{RaichevMISO,Raichev2008,RevModPhysVanya}

\begin{equation}\label{eq:twoband}
	\Delta \rho=\frac{2\tau}{\tau_\mathrm{inter}} \rho_\mathrm{0} \delta_\mathrm{1}\delta_\mathrm{2}  \cos(2\pi\Delta\nu/g).
\end{equation}
Here $\rho_\mathrm{0}$ is the Drude resistivity, $\delta_\mathrm{1,2}=\exp(-\pi/\omega_\mathrm{c}\tau_\mathrm{q1,2})$ are the Dingle factors of the two subbands labeled by indices 1 and 2 and expressed in terms of the cyclotron frequency, $\omega_c$, and the quantum scattering times $\tau_\mathrm{q1}$ and $\tau_\mathrm{q2}$, $g$ is the subband degeneracy, $\Delta \nu=\nu_\mathrm{2}-\nu_\mathrm{1}$ is the difference in filling factors, $\nu_\mathrm{1,2}=n_\mathrm{1,2} h/B e$, of the  subbands, and $\tau_\mathrm{inter}$ is the interband scattering time.  \DAB{Note, while initially derived for 2DES with parabolic spectrum, Eq. (1) becomes generally applicable when expressed in terms of the filling factors. Indeed, $\Delta \nu/g$ universally determines the condition where the LLs in both subbands are aligned. In addition, we mention that the conditions of our experiments actually correspond to high filling factors and low $T$ where the effects of non-parabolicity and associated non-equidistant LL spectrum are negligible.}

\begin{figure*}[ht!]
	\centering
	\includegraphics[width=1\textwidth]{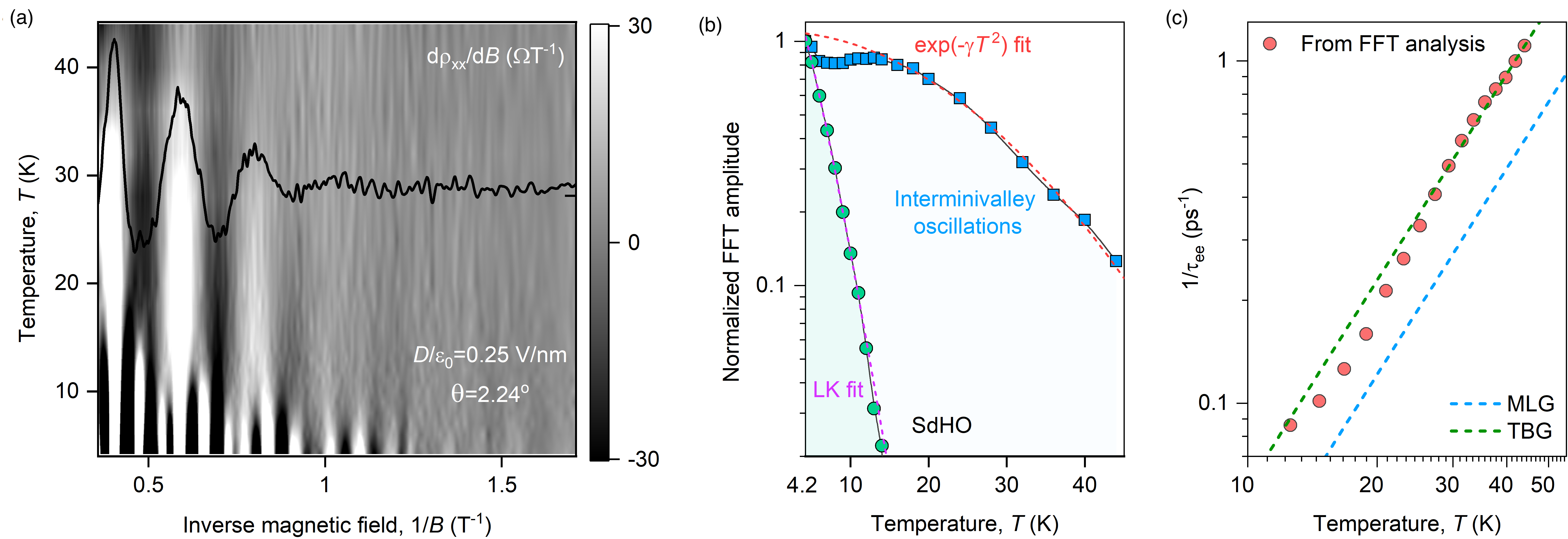}
\caption{Temperature dependence of the interminivalley magnetooscillations in SA-TBG. (a) $\mathrm{d}\rho_\mathrm{xx}/\mathrm{d}B$ mapped against $T$ and $1/B$ at $n=2.76\times10^{12}\mathrm{cm}^{-2}$ and $\frac{D}{\varepsilon_\mathrm{0}}=0.25~$V/nm. (b) FFT amplitude of the SdHO and MP oscillations as a function of $T$. Red dashed line: Fit of the high-$T$ region with $~\mathrm{exp}(-\gamma T^2)$, where $\gamma=11.5\times10^{-4}~\mathrm{K}^{-2}$. Purple dashed line: LK law fit of SdHO. (c) Experimentally-derived e-e scattering rate ($1/\tau_{\mathrm{ee}}$) versus $T$ (red dots). Blue and green dashed lines are the $1/\tau_{\mathrm{ee}}(T)$ dependence for MLG taken from Ref.~\cite{TIMO} and the results (green dashed curve) of its renormalization accounting for reduced $v_\mathrm{F}$ and eight-fold degeneracy in SA-TBG at small fillings.}
    \label{fig:Fig3}
\end{figure*}

To validate the interpretation of the observed high-$T$ magnetooscillations in SA-TBG in the context of MISO physics, we plot the experimentally determined $B_\mathrm{0}$ (from Fig.~\ref{fig:Fig2}b) as a function of $\Delta n$ in the inset of Fig.~\ref{fig:Fig2}b. This difference in carrier density, $\Delta n$, was obtained by a simple electrostatics argument that accounts for the partial screening of the applied field by the graphene layers~\cite{VolodyaScreening,SanchezPRL,berdyugin2020} (Supplemental Material~\cite{Supplementary}). Additionally, for some $D$, we  also verified the aforementioned $\Delta n $ values by FFT analysis at liquid helium $T$ as well, where the beating of SdHO can be used to determine $\Delta n$. For all our devices, the obtained $B_\mathrm{0}(\Delta n)$ dependence was found to be linear over a wide range of $\Delta n$ and accurately followed  the functional form $B_\mathrm{0}=h \Delta n/4 e$, where $4$ represents the degeneracy of each minivalley. This substantiates the interpretation of the observed oscillations in terms of MISO, where minivalleys now take on the role of the subbands. 

Another important characteristic of MISO is its fragile phase with respect to dc bias~\cite{Bykov2008,HIRO_MISO,Wiedmann2011,Drichko2020}. In the presence of magnetic field, an electric current, $I_\mathrm{dc}$, of high density generates a substantial Hall field perpendicular to the current flow that initiates additional impurity-assisted tunneling of electrons between the tilted LLs~\cite{Reno,Glazman,ZudovGraphene}. The probability of such tunneling events oscillates with magnetic field and is maximized when the Hall voltage drop across the cyclotron diameter matches an integer multiple of the cyclotron energy~\cite{RevModPhysVanya}. This leads to the modification of the resonant condition for MISO which manifests itself in multiple phase reversals upon ramping $I_\mathrm{dc}$. This interesting behaviour was also found in our SA-TBG devices, which exhibited the aforementioned phase flips with respect to $I_\mathrm{dc}$ (Supplemental Material and Fig. S2~\cite{Supplementary}, further supporting the origin of the observed oscillations.

Moreover, unlike SdHO, which also emerge as a result of the Landau quantization, the intersubband oscillations are not sensitive to the smearing of the Fermi distribution, and therefore are damped only through the broadening of LLs, parameterized via the Dingle factors in Eq.~\ref{eq:twoband}~\cite{HIRO_MISO,RevModPhysVanya}. Our data reveals this expected behaviour too: namely, the FFT amplitude of the interminivalley oscillations features a slow $\exp(-\gamma T^2)$ decay ($\gamma=11.5\times 10^{-4}~$K$^{-2}$), as compared to the relatively fast SdHO thermal damping governed by the conventional Lifshitz-Kosevich (LK) law (dashed purple line in Fig.~\ref{fig:Fig3}b). This behaviour is also consistent with the robustness of the interminivalley oscillations to heating induced by large $I_{\mathrm{dc}}$ (See Supplemental Material and Fig.~S2 for details~\cite{Supplementary}).  

The observed high-$T$ magnetooscillations provide a convenient tool to estimate the relative ratio between inter- and intraminivalley scattering rates in SA-TBG by fitting them with Eq.~\ref{eq:twoband}. From the exponential damping of the oscillations' amplitude with decreasing $B$, one can extract the quantum scattering time, while $\rho_\mathrm{0}$ can be obtained from the zero-$B$ data leaving $\tau/\tau_\mathrm{inter}$ as the only fitting parameter. We have performed such an analysis for our smallest angle device and, from the data shown in Fig.~\ref{fig:Fig1}f, found that at $T=16~$K, $\tau$ and $\tau_\mathrm{inter}$ are comparable, indicating the significance of interminivalley scattering processes at small $\theta$ (see Supplemental Material and Fig.~S3 for details~\cite{Supplementary}). \DAB{We also note a drop in the oscillations' amplitude with increasing $\theta$. This indicates the suppression of  interminivalley scattering at larger twist angles likely because a larger momentum gain, $\sim \Delta k$, is required to initiate such transitions. However, more accurate comparison on samples with identical quality is needed to verify such conclusion. }

The interminivalley oscillations also provide an additional insight into the electronic properties of SA-TBG: the observed $\exp(-\gamma T^2)$ behaviour of the FFT amplitude (Fig.~\ref{fig:Fig3}b) suggests LL broadening induced by e-e scattering~\cite{RaichevMISO,RevModPhysVanya} and thus, its rate, $1/\tau_\mathrm{ee}$, can be conveniently estimated via an analysis of the oscillations' thermal damping~\cite{Vitkalov_ee}. Assuming that this thermal damping is solely encoded in $\delta_\mathrm{1}\delta_\mathrm{2}$ through the temperature dependence of the quantum scattering times (see Eq.~\ref{eq:twoband}), and that these are identical in both minivalleys (a reasonable assumption when $\Delta n\ll n$)~\cite{RaichevMISO}, one obtains the $T-$dependent amplitude of the interminivalley oscillations: $\delta_\mathrm{1} \delta_\mathrm{2}=e^{-2 \pi/\omega_\mathrm{c}\tau_q(T)}$. Since  $\tau^{-1}_\mathrm{q}(T)=\tau^{-1}_\mathrm{0}+\tau^{-1}_\mathrm{ee}(T)$, where $\tau_\mathrm{0}$ is the $T-$independent elastic quantum scattering time, one can extract $\tau^{-1}_\mathrm{ee}$ from the FFT magnitude of the interminivalley oscillations. 
Figure~\ref{fig:Fig3}c shows the results of such an analysis and plots the $\tau^{-1}_\mathrm{ee}(T)$ dependence. For $\theta=2.24\degree$, we find that the obtained estimates exceed the e-e scattering rate in MLG (blue dashed curve in Fig.~\ref{fig:Fig3}c) at identical $n$~\cite{TIMO,polini2014quasiparticle,kumar2017superball}.  We attribute this enhancement to the reduced $v_\mathrm{F}=0.75 v_\mathrm{0}$ in the SA-TBG of this $\theta$ as compared to that of MLG, $v_\mathrm{0}=10^6~$m/s. Indeed, by renormalizing the $1/\tau_\mathrm{ee}(T)$ dependence for MLG from Fig.~\ref{fig:Fig3}c by the  ratio of the Fermi velocities in these systems and accounting for the two fold increase in the degeneracy of SA-TBG as compared to MLG, we obtain the scattering rate for SA-TBG close to that found experimentally.  

Our experiments also raise important questions about scattering processes in twisted moir\'e systems. The observed high-$T$ oscillations indicate the presence of some scattering mechanism(s) enabling electrons to gain enough momentum ($\sim \Delta k$) to escape from their minivalley and scatter to another one ($k_\mathrm{m} \leftrightarrow k'_\mathrm{m}$). This, in turn, may imply the presence of scatterers with a spatial scale of the order of $1/\Delta k\sim\lambda_\mathrm{m}$, where $\lambda_\mathrm{m}$ is the superlattice period. A possible candidate is twist angle disorder, regularly observed in devices fabricated by the methods used here~\cite{Twistdisorder}. An alternative scenario involves acoustic phonon-assisted processes~\cite{RaichevMISOphonons,RoshMP,MarkMP}. However, in our data, the interminivalley oscillations are present even at liquid helium $T$ (Fig.~\ref{fig:Fig3}b), at which the allowed phase space for phonon momenta is not sufficient to ensure the momentum mismatch $\Delta k$. At $T=4.2~$K, phonons with momenta $q<k_\mathrm{B}T/\hbar s \approx ~4\times10^7$~m$^\mathrm{-1}$ are populated (where $s\approx20$~km/s is the characteristic speed of sound in graphene),
i.e., those having momenta over an order of magnitude smaller than $\Delta k$ at the studied twist angles. The only phonon branch that at such low $T$ can be populated up to the required momenta, is the breathing mode~\cite{Balandin}; however, little is known on its impact on SA-TBG resistivity~\cite{RayTBG}. 

To conclude, we have observed high-$T$ magnetooscillations in SA-TBG when a finite displacement field is applied across the graphene layers. Although similarly periodic in $1/B$, these oscillations show a clearly distinct temperature and dc current dependence from SdHO and are controlled by the difference in the minivalleys' filling factors.  Drawing a parallel with MISO, we have shown that the observed oscillations originate from interminivalley scattering, allowed by the reduced size of the mini-Brillouin zone in SA-TBG. By analyzing the amplitude of these high-$T$ oscillations, we estimated the relative ratio between interminivalley and intraminivalley scattering times, $\tau/\tau_{\mathrm{inter}}$, which we found to be of similar order in the $\theta=1.65\degree$ device. Finally, from the temperature dependence of the oscillations, we obtained information on $\tau^{-1}_{\mathrm{ee}}$, which we found to exceed that of MLG due to a reduced $v_\mathrm{F}$ in this system. Our study points to the presence of a scattering mechanism(s) of unknown nature with large momentum transfer and highlights the importance of interminivalley momentum relaxation in the resistivity of twisted moir\'e systems~\cite{Chung,Polshyn2019,Rickhaus_intervalley,Rosh_BZ,Wallbank2019} that has to be accounted for in future studies.

 \section*{Acknowledgments}
This work was supported by AFOSR grant FA9550-16-1-0382 and the Gordon and Betty Moore Foundation's EPiQS Initiative through Grant GBMF9643 to P.J.-H. D.A.B. acknowledges the support from MIT Pappalardo Fellowship. I.Y.P acknowledges support from the MIT undergraduate research opportunities program and the Johnson \& Johnson research scholars program. I.A.D. acknowledges support from the German Research Foundation under DFG projects GA501/17-1 (SPP 2244) and DM1/5-1. Growth of hBN crystals was supported by the Elemental Strategy Initiative conducted by the MEXT, Japan, Grant JPMXP0112101001, JSPS KAKENHI Grant JP20H00354. We thank Laurence Eaves, Mark Greenaway, Leonid Levitov, Michael Zudov, Volodya Fal'ko, Roshan Krishna Kumar, Vasili Perebeinos, Sergio C. de la Barrera, Alexander A. Zibrov, Alexey Berdyugin and Mallika Randeria for fruitful discussions.

\noindent\rule{6cm}{0.4pt}

*Correspondence to: \\ bandurin@mit.edu, pjarillo@mit.edu.

%


\newpage

\setcounter{figure}{0}
\renewcommand{\thesection}{}
\renewcommand{\thesubsection}{S\arabic{subsection}}
\renewcommand{\theequation} {S\arabic{equation}}
\renewcommand{\thefigure} {S\arabic{figure}}
\renewcommand{\thetable} {S\arabic{table}}
\begin{widetext}
\newpage

\section{\textbf{Supplemental Material}}

\subsection{\textbf{Supplementary Section 1. Device fabrication} }

Our devices consisted of hBN-encapsulated twisted bilayer graphene, which we fabricated using a combination of cut-and-stack~\cite{Tutuc_TBG, lasercut} and hot release~\cite{Hot-transfer_NComm} methods. Monolayer graphene, few-layer graphite, and 30-80\,nm-thick hBN crystals were mechanically exfoliated on a Si/SiO$_2$ substrate, and sizable, uniform flakes were selected using optical contrast. Then, using a homemade transfer system with $\mu$m-accuracy and a polycarbonate (PC) membrane stretched over a small (8\,mm$\times$8\,mm$\times$4\,mm) polydimethylsiloxane (PDMS) polymer block on a glass slide, we assembled hBN and graphite stacks on a Si/SiO$_2$ wafer. To minimize strain on the hBN, we picked up at 50-70\,\degree C, when the membrane was minimally sticky enough to allow for a clean pickup. The graphite was picked up at room temperature, and then the entire stack was ``ironed'' and then released on a clean Si/SiO$_2$ wafer at high temperatures ($160~-~170$\,\degree C). After removing the polymer membrane, we annealed the hBN and graphite stack at 350\,\degree C for 3 hours while flowing argon and hydrogen in order to ensure the removal of any residues. We then assembled the hBN and twisted bilayer graphene stack using a ``cut-and-stack'' method described previously~\cite{Tutuc_TBG,YuanPRL}. After picking up the top hBN and twisted graphene, we ``ironed'' the entire stack at room temperature. The three-layer stack was then released onto the previously fabricated and cleaned bottom hBN and graphite gate at roughly 160\,C. After this point, we avoided heating the stack to reduce the possibility of twist angle relaxation. The resulting heterostructure is shown in Fig.~\ref{fig:FigSFab}a. The final stack was inspected using dark-field microscopy (Fig.~\ref{fig:FigSFab}b) and atomic force microscopy (AFM), and bubble- and blister-free areas were selected to use for Hall bars (Fig.~\ref{fig:FigSFab}c).

To fabricate the devices, we covered the heterostructures by a protective polymethyl-methacrylate (PMMA) resist and used electron beam lithography (EBL) to define contact regions (Fig.~\ref{fig:FigSFab}d). We then performed a mild O$_2$ plasma cleaning before using reactive ion etching (RIE) with a plasma generated from  CHF$_3$ and O$_2$ gases to selectively etch away the hBN in the parts of the heterostructure unprotected by the lithographic mask~\cite{MosheNatPhys}. 3\,nm chromium and 50-70\,nm gold was then evaporated into the contact regions via thermal evaporation at high vacuum (Fig.~\ref{fig:FigSFab}e). We repeat the same EBL and thermal evaporation procedures to define a metallic top gate (3\,nm chromium and 30-40\,nm gold). Finally, we repeat the same EBL and RIE procedures to define the final Hall bar geometry, using, in this case, a plasma generated by Ar, O$_2$ and CHF$_3$ gases.

\begin{figure*}[ht!]
	\centering
	\includegraphics[width=0.7\textwidth]{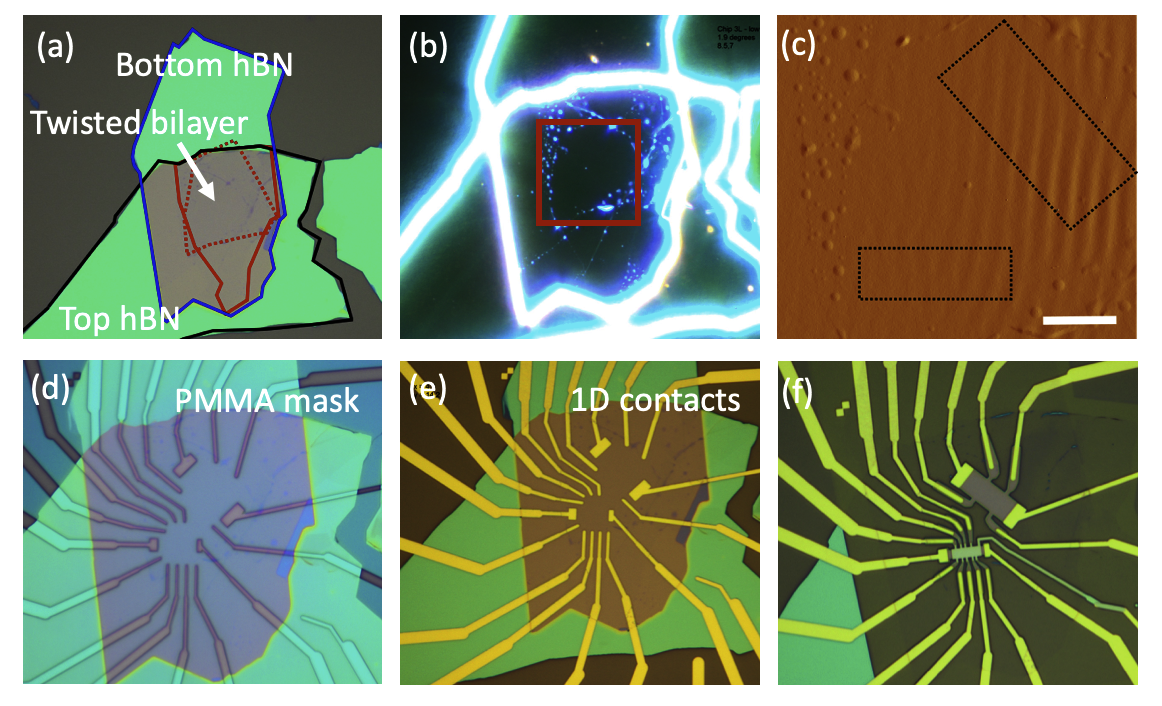}
	\caption{\textbf{Device fabrication} \textbf{a,} Optical photograph of a twisted bilayer graphene encapsulated between two slabs of hBN. \textbf{b}, Dark field image of the obtained heterostructure with blister-free area that was further examined by AFM (red). \textbf{c,} AFM topography of regions used for Hall bars. Ripples are an artifact of the AFM scanning and are not physical. White bar is 5\,$\mu$m. Area used for Hall bar indicated by dashed black rectangle. \textbf{d,} Protective PMMA mask used to define contact areas via RIE. \textbf{e,} Gold contacts are evaporated on to graphene after RIE. \textbf{f,} Optical photograph of the finalized device. }
\label{fig:FigSFab}
\end{figure*}

\newpage
\subsection{\textbf{Supplementary Section 2. Angle determination}}

Because of the relatively large $\theta$ for two of our devices, we were limited by the robustness of the gate dielectric and could not reach full-filling of the moir\'e bands, which is the standard approach of determining twist angle. To circumvent this problem, we performed further magnetoresistance measurements and, for one of the devices, resolved well-defined Brown-Zak oscillations, whose fundamental frequency provided an accurate tool for determining the superlattice period~\cite{Rosh_BZ} and therefore the corresponding twist angle $2.24\degree$. For the second device, where Brown-Zak oscillations were not detected, we performed a detailed measurements of the effective mass of charge carriers, $m$, using $T$-induced damping of the SdHO and compared it with that of the $2.24\degree$ SA-TBG. This comparison pointed to $10-20$\% lighter charge carriers in the $2.24\degree$ SA-TBG at the same $n$ which provided a rough estimate for the twist angle $2.3\degree$ using a theoretical framework~\cite{bistritzer2011}. Lastly, from the superlattice-induced insulating states at full-filling ($n_s$) readily observed in the $\rho_\mathrm{xx}(n)$ dependencies of our smallest-angle SA-TBG (Fig.~\ref{fig:Fig1}c) and the relation $n_s = 4\times\frac{8 \sin^2(\theta/2)}{\sqrt{3}a^2}$, we determine $\theta\approx1.65\degree$. 

 \newpage
 \subsection{\textbf{Supplementary Section 3.  Connection between the applied gate voltages and density imbalance}}
 
As discussed in the main text, to validate the interpretation of the observed high-$T$ magnetooscillations in SA-TBG in the context of MISO physics, one has to know the difference in carrier density $\Delta n$ between the two minivalleys for a given displacement field $D$ applied between the graphene layers. This difference can obtained by a simple electrostatics consideration that accounts for the electrostatic screening  effect~\cite{VolodyaScreening,SanchezPRL,berdyugin2020}:
 \begin{align}\label{eq:screening}
     \frac{ed}{\epsilon_0 \epsilon}\left( \epsilon_0 D -  \frac{1+\epsilon}{4} (n_1 - n_2)e\right) &= \frac{h v_\mathrm{0} }{2 \sqrt{\pi}} \left(s_1 \sqrt{|n_1|} - s_2 \sqrt{|n_2|}\right), \\
     n &= n_1 + n_2,
 \end{align}
 where  $d$ is the distance separating the graphene layers,  $e$  is the electron charge, $\epsilon$ is the dielectric constant for twisted bilayer~\cite{SanchezPRL,berdyugin2020}, $v_\mathrm{0}$ is the Dirac velocity and $\epsilon_0$ is the vacuum permittivity; the band indices are given by $s_\mathrm{i}=n_\mathrm{i}/|n_\mathrm{i}|$. To find $\Delta n$ for each value of $n$ and $D$ used in the experiment, the two equations are solved simultaneously using $\epsilon = 2.7$ known from previous experiments~\cite{VolodyaScreening,SanchezPRL,berdyugin2020,VolodyaBeating}. 
 
\newpage

\subsection{\textbf{Supplementary Section 4. Interminivalley oscillations under strong dc bias}}

To further explore the properties of the observed high-$T$ interminivalley oscillations, we have performed the measurements of the differential resistance, $r=\mathrm{d}V/\mathrm{d}I$, as a function of DC current $I_\mathrm{dc}$ and magnetic field, $B$, in our $\theta=2.3\degree$ device. Figure~\ref{fig:FigPhase}b shows the $r(B)$ dependence along with its derivative $\mathrm{d}r/\mathrm{d}B$ (red) measured at $I_\mathrm{dc}=0~\mu$A (black) and $T=20$~K. The data resembles that shown in Fig.~\ref{fig:Fig2}a of the main text. The differentiation improves the visibility of the oscillations by removing the smooth non-oscillating $I_\mathrm{dc}$-dependent background. With increasing $I_\mathrm{dc}$, the amplitude of the oscillations remains practically unaffected whereas the phase changes for certain values of $I_\mathrm{dc}$ as apparent from the checkboard pattern of the $\mathrm{d}r/\mathrm{d}B(I_\mathrm{dc},B)$ map shown in Fig.~\ref{fig:FigPhase}a. This pattern persists up to the highest currents of $I_\mathrm{dc}\approx75~\mu$A applied in our experiments with no apparent decay of the oscillations' amplitude. For comparison, SdHO are washed out with the application of only $I_\mathrm{dc}\sim10~\mu$A if similar measurements are performed at $T=4.2~$K. These observations clearly indicate that the observed high-$T$ magnetooscillations are entirely different than SdHO. 

Qualitatively, in the presence of magnetic field, an electric current of high density generates a substantial Hall field perpendicular to the current flow. In sufficiently clean systems, this field initiates scattering-assisted transitions of electrons between the tilted Landau levels~\cite{Reno,Glazman,ZudovGraphene}. The probability of these transitions is maximized when the Hall voltage drop across the cyclotron diameter matches an integer multiple of the cyclotron energy~\cite{RevModPhysVanya} giving rise to peculiar magnetooscillations sensitive to the value of $I_\mathrm{dc}$. The observed phase flip and the corresponding checkboard pattern shown in Fig.~\ref{fig:FigPhase}(a) demonstrates the modification of the intersubband scattering in SA-TBG by Hall voltage-induced transitions~\cite{Bykov2008,HIRO_MISO,Wiedmann2011,Drichko2020}.

\begin{figure*}[ht!]
	\centering
	\includegraphics[width=0.45\textwidth]{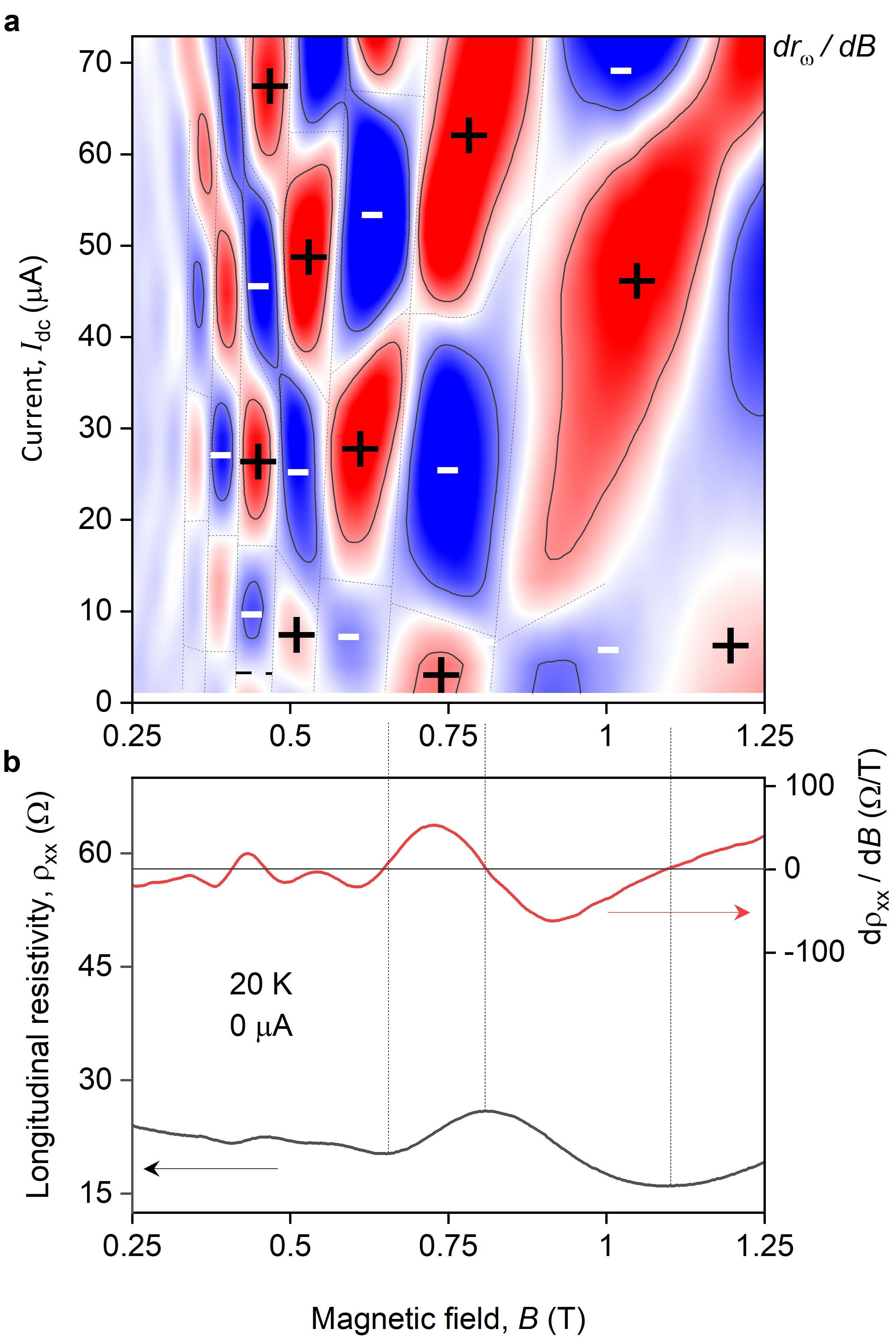}
	\caption{\textbf{Interminivalley oscillations in the presence of strong dc field. } \textbf{a,} The derivative of the differential resistance with respect to magnetic field, $\mathrm{d}r/\mathrm{d}B$, mapped against $B$ and $I_\mathrm{dc}$. \textbf{b,} The interminivalley oscillations observed in $r(B)$ (black) and its derivative $\mathrm{d}r/\mathrm{d}B$ at $I_\mathrm{dc}=0~\mu\mathrm{A}$. $\theta=2.3\degree$, $T=20$~K, $n=1.5\times 10^{12}$~cm$^{-2}$. }  \label{fig:FigPhase}
\end{figure*}

\newpage

\subsection{Supplementary Section 5: Estimating $\tau / \tau_{\mathrm{inter}}$ in SA-TBG}

The observed interminivalley oscillations reported in the main text provide a tool for characterizing momentum relaxation processes in SA-TBG. Namely, they enable estimating the ratio of the interminivalley scattering to the total scattering rate. Figure~\ref{fig:FigStau} plots the amplitude of the interminivalley oscillations, $\rho_\mathrm{xx}$, as a function of inverse magnetic field, $1/B$, obtained by subtracting a smooth, non-oscillating background from the data reported in Fig.~1f of the main text. The oscillations feature a  damped sinusoidal dependence on $1/B$ described by Eq.~\ref{eq:twoband} of the main text. In the assumption of equal Dingle factors in both minivalleys $\delta_\mathrm{1}=\delta_\mathrm{2}$ (a fair assumption as long as $\delta n\ll n$), this equation becomes:
\begin{equation}\label{eq:twobandS4}
    \Delta \rho=\frac{2 \tau}{\tau_\mathrm{inter}} \rho_\mathrm{0} e^{-2\pi m/eB \tau_q} \cos{(2 \pi B_\mathrm{0}/B)},
\end{equation}
where $\rho_\mathrm{0}$ is the resistivity at $B=0$, $m$ is the effective mass of the charge carriers, $\tau_\mathrm{q}$ is the the quantum scattering time, and $\tau/\tau_\mathrm{inter}$ is the ratio between the intraminivalley and interminivalley scattering times. Using $m$, obtained from the analysis of the SdHO thermal damping, $\tau_\mathrm{q}$ from the Dingle plot shown in the inset of Fig.~\ref{fig:FigStau} and $\rho_\mathrm{0}$ from the zero-$B$ measurements, we plot the calculated $\Delta \rho(1/B)$ dependence which shows good agreement with experiment for $\tau/\tau_\mathrm{inter}=3/4$. For comparison, we also illustrate the case when interminivalley scattering is weaker,  $\tau/\tau_\mathrm{inter}=1/3$, which shows a reduced oscillations amplitude. These findings indicate the importance of interminivalley scattering in SA-TBG. 

\begin{figure*}[ht!]
	\centering
	\includegraphics[width=0.5\textwidth]{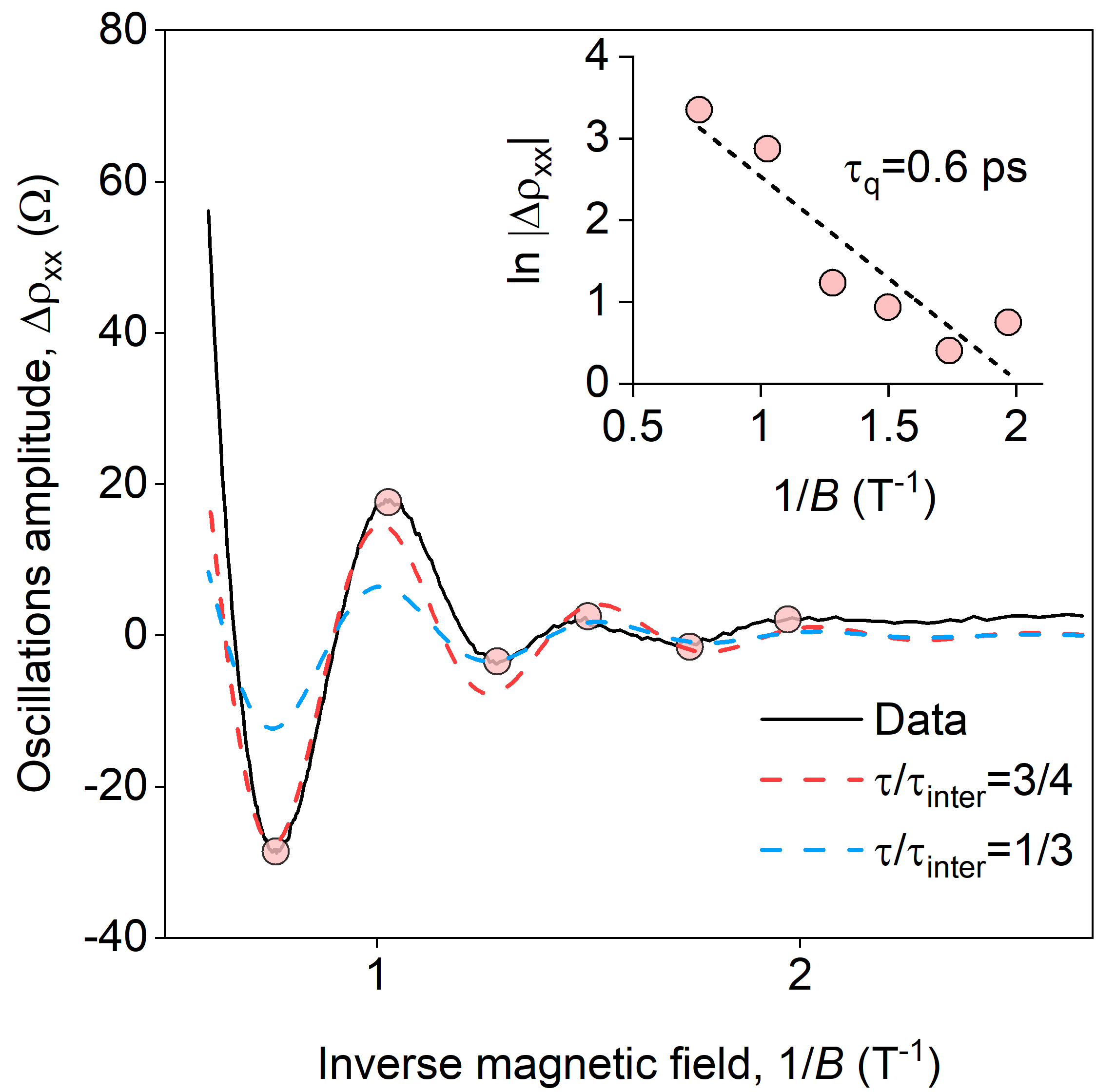}
	\caption{\textbf{Estimating $\tau / \tau_{\mathrm{inter}}$. } Solid: Oscillation amplitude, $\Delta \rho_\mathrm{xx}$, as a function of $1/B$, obtained by subtracting a smooth non-oscillating background from the data in Fig. 1f (red curve). Red and blue dashed lines: intersubband resistivity oscillations calculated from Eq. (1) for $\tau/\tau_\mathrm{inter}=3/4$ and  $\tau/\tau_\mathrm{inter}=1/3$ and experimentally determined $\rho_\mathrm{D}=122~\Omega$, $m=0.044m_\mathrm{e}$, and $\tau_\mathrm{q}=0.6~$ps. Inset: Dingle plot for the interminivalley oscillations in $\theta=1.65\degree$ SA-TBG along with the best fit to an exponential decay function (black dashed line) yielding $\tau_\mathrm{q}=0.6~$ps.} 
	\label{fig:FigStau}
\end{figure*}

\newpage

\end{widetext}
%

\begin{thebibliography}{53}%
	\makeatletter
	\providecommand \@ifxundefined [1]{%
		\@ifx{#1\undefined}
	}%
	\providecommand \@ifnum [1]{%
		\ifnum #1\expandafter \@firstoftwo
		\else \expandafter \@secondoftwo
		\fi
	}%
	\providecommand \@ifx [1]{%
		\ifx #1\expandafter \@firstoftwo
		\else \expandafter \@secondoftwo
		\fi
	}%
	\providecommand \natexlab [1]{#1}%
	\providecommand \enquote  [1]{``#1''}%
	\providecommand \bibnamefont  [1]{#1}%
	\providecommand \bibfnamefont [1]{#1}%
	\providecommand \citenamefont [1]{#1}%
	\providecommand \href@noop [0]{\@secondoftwo}%
	\providecommand \href [0]{\begingroup \@sanitize@url \@href}%
	\providecommand \@href[1]{\@@startlink{#1}\@@href}%
	\providecommand \@@href[1]{\endgroup#1\@@endlink}%
	\providecommand \@sanitize@url [0]{\catcode `\\12\catcode `\$12\catcode
		`\&12\catcode `\#12\catcode `\^12\catcode `\_12\catcode `\%12\relax}%
	\providecommand \@@startlink[1]{}%
	\providecommand \@@endlink[0]{}%
	\providecommand \url  [0]{\begingroup\@sanitize@url \@url }%
	\providecommand \@url [1]{\endgroup\@href {#1}{\urlprefix }}%
	\providecommand \urlprefix  [0]{URL }%
	\providecommand \Eprint [0]{\href }%
	\providecommand \doibase [0]{http://dx.doi.org/}%
	\providecommand \selectlanguage [0]{\@gobble}%
	\providecommand \bibinfo  [0]{\@secondoftwo}%
	\providecommand \bibfield  [0]{\@secondoftwo}%
	\providecommand \translation [1]{[#1]}%
	\providecommand \BibitemOpen [0]{}%
	\providecommand \bibitemStop [0]{}%
	\providecommand \bibitemNoStop [0]{.\EOS\space}%
	\providecommand \EOS [0]{\spacefactor3000\relax}%
	\providecommand \BibitemShut  [1]{\csname bibitem#1\endcsname}%
	\let\auto@bib@innerbib\@empty
	\bibitem [{\citenamefont {Cao}\ \emph {et~al.}(2018{\natexlab{a}})\citenamefont
		{Cao}, \citenamefont {Fatemi}, \citenamefont {Fang}, \citenamefont
		{Watanabe}, \citenamefont {Taniguchi}, \citenamefont {Kaxiras},\ and\
		\citenamefont {Jarillo-Herrero}}]{Yuan_SC}%
	\BibitemOpen
	\bibfield  {author} {\bibinfo {author} {\bibfnamefont {Y.}~\bibnamefont
			{Cao}}, \bibinfo {author} {\bibfnamefont {V.}~\bibnamefont {Fatemi}},
		\bibinfo {author} {\bibfnamefont {S.}~\bibnamefont {Fang}}, \bibinfo {author}
		{\bibfnamefont {K.}~\bibnamefont {Watanabe}}, \bibinfo {author}
		{\bibfnamefont {T.}~\bibnamefont {Taniguchi}}, \bibinfo {author}
		{\bibfnamefont {E.}~\bibnamefont {Kaxiras}}, \ and\ \bibinfo {author}
		{\bibfnamefont {P.}~\bibnamefont {Jarillo-Herrero}},\ }\href {\doibase
		10.1038/nature26160} {\bibfield  {journal} {\bibinfo  {journal} {Nature}\
		}\textbf {\bibinfo {volume} {556}},\ \bibinfo {pages} {43} (\bibinfo {year}
		{2018}{\natexlab{a}})}\BibitemShut {NoStop}%
	\bibitem [{\citenamefont {Yankowitz}\ \emph {et~al.}(2019)\citenamefont
		{Yankowitz}, \citenamefont {Chen}, \citenamefont {Polshyn}, \citenamefont
		{Zhang}, \citenamefont {Watanabe}, \citenamefont {Taniguchi}, \citenamefont
		{Graf}, \citenamefont {Young},\ and\ \citenamefont {Dean}}]{Yankowitz}%
	\BibitemOpen
	\bibfield  {author} {\bibinfo {author} {\bibfnamefont {M.}~\bibnamefont
			{Yankowitz}}, \bibinfo {author} {\bibfnamefont {S.}~\bibnamefont {Chen}},
		\bibinfo {author} {\bibfnamefont {H.}~\bibnamefont {Polshyn}}, \bibinfo
		{author} {\bibfnamefont {Y.}~\bibnamefont {Zhang}}, \bibinfo {author}
		{\bibfnamefont {K.}~\bibnamefont {Watanabe}}, \bibinfo {author}
		{\bibfnamefont {T.}~\bibnamefont {Taniguchi}}, \bibinfo {author}
		{\bibfnamefont {D.}~\bibnamefont {Graf}}, \bibinfo {author} {\bibfnamefont
			{A.~F.}\ \bibnamefont {Young}}, \ and\ \bibinfo {author} {\bibfnamefont
			{C.~R.}\ \bibnamefont {Dean}},\ }\href {\doibase 10.1126/science.aav1910}
	{\bibfield  {journal} {\bibinfo  {journal} {Science}\ }\textbf {\bibinfo
			{volume} {363}},\ \bibinfo {pages} {1059} (\bibinfo {year}
		{2019})}\BibitemShut {NoStop}%
	\bibitem [{\citenamefont {Cao}\ \emph {et~al.}(2018{\natexlab{b}})\citenamefont
		{Cao}, \citenamefont {Fatemi}, \citenamefont {Demir}, \citenamefont {Fang},
		\citenamefont {Tomarken}, \citenamefont {Luo}, \citenamefont
		{Sanchez-Yamagishi}, \citenamefont {Watanabe}, \citenamefont {Taniguchi},
		\citenamefont {Kaxiras}, \citenamefont {Ashoori},\ and\ \citenamefont
		{Jarillo-Herrero}}]{Yuan_Insulators}%
	\BibitemOpen
	\bibfield  {author} {\bibinfo {author} {\bibfnamefont {Y.}~\bibnamefont
			{Cao}}, \bibinfo {author} {\bibfnamefont {V.}~\bibnamefont {Fatemi}},
		\bibinfo {author} {\bibfnamefont {A.}~\bibnamefont {Demir}}, \bibinfo
		{author} {\bibfnamefont {S.}~\bibnamefont {Fang}}, \bibinfo {author}
		{\bibfnamefont {S.~L.}\ \bibnamefont {Tomarken}}, \bibinfo {author}
		{\bibfnamefont {J.~Y.}\ \bibnamefont {Luo}}, \bibinfo {author} {\bibfnamefont
			{J.~D.}\ \bibnamefont {Sanchez-Yamagishi}}, \bibinfo {author} {\bibfnamefont
			{K.}~\bibnamefont {Watanabe}}, \bibinfo {author} {\bibfnamefont
			{T.}~\bibnamefont {Taniguchi}}, \bibinfo {author} {\bibfnamefont
			{E.}~\bibnamefont {Kaxiras}}, \bibinfo {author} {\bibfnamefont {R.~C.}\
			\bibnamefont {Ashoori}}, \ and\ \bibinfo {author} {\bibfnamefont
			{P.}~\bibnamefont {Jarillo-Herrero}},\ }\href {\doibase 10.1038/nature26154}
	{\bibfield  {journal} {\bibinfo  {journal} {Nature}\ }\textbf {\bibinfo
			{volume} {556}},\ \bibinfo {pages} {80} (\bibinfo {year}
		{2018}{\natexlab{b}})}\BibitemShut {NoStop}%
	\bibitem [{\citenamefont {{Sharpe}}\ \emph {et~al.}(2019)\citenamefont
		{{Sharpe}}, \citenamefont {{Fox}}, \citenamefont {{Barnard}}, \citenamefont
		{{Finney}}, \citenamefont {{Watanabe}}, \citenamefont {{Taniguchi}},
		\citenamefont {{Kastner}},\ and\ \citenamefont
		{{Goldhaber-Gordon}}}]{sharpe2019}%
	\BibitemOpen
	\bibfield  {author} {\bibinfo {author} {\bibfnamefont {A.~L.}\ \bibnamefont
			{{Sharpe}}}, \bibinfo {author} {\bibfnamefont {E.~J.}\ \bibnamefont {{Fox}}},
		\bibinfo {author} {\bibfnamefont {A.~W.}\ \bibnamefont {{Barnard}}}, \bibinfo
		{author} {\bibfnamefont {J.}~\bibnamefont {{Finney}}}, \bibinfo {author}
		{\bibfnamefont {K.}~\bibnamefont {{Watanabe}}}, \bibinfo {author}
		{\bibfnamefont {T.}~\bibnamefont {{Taniguchi}}}, \bibinfo {author}
		{\bibfnamefont {M.~A.}\ \bibnamefont {{Kastner}}}, \ and\ \bibinfo {author}
		{\bibfnamefont {D.}~\bibnamefont {{Goldhaber-Gordon}}},\ }\href@noop {}
	{\bibfield  {journal} {\bibinfo  {journal} {Science}\ }\textbf {\bibinfo
			{volume} {365}},\ \bibinfo {pages} {605} (\bibinfo {year}
		{2019})}\BibitemShut {NoStop}%
	\bibitem [{\citenamefont {{Serlin}}\ \emph {et~al.}(2020)\citenamefont
		{{Serlin}}, \citenamefont {{Tschirhart}}, \citenamefont {{Polshyn}},
		\citenamefont {{Zhang}}, \citenamefont {{Zhu}}, \citenamefont {{Watanabe}},
		\citenamefont {{Taniguchi}}, \citenamefont {{Balents}},\ and\ \citenamefont
		{{Young}}}]{serlin2020}%
	\BibitemOpen
	\bibfield  {author} {\bibinfo {author} {\bibfnamefont {M.}~\bibnamefont
			{{Serlin}}}, \bibinfo {author} {\bibfnamefont {C.~L.}\ \bibnamefont
			{{Tschirhart}}}, \bibinfo {author} {\bibfnamefont {H.}~\bibnamefont
			{{Polshyn}}}, \bibinfo {author} {\bibfnamefont {Y.}~\bibnamefont {{Zhang}}},
		\bibinfo {author} {\bibfnamefont {J.}~\bibnamefont {{Zhu}}}, \bibinfo
		{author} {\bibfnamefont {K.}~\bibnamefont {{Watanabe}}}, \bibinfo {author}
		{\bibfnamefont {T.}~\bibnamefont {{Taniguchi}}}, \bibinfo {author}
		{\bibfnamefont {L.}~\bibnamefont {{Balents}}}, \ and\ \bibinfo {author}
		{\bibfnamefont {A.~F.}\ \bibnamefont {{Young}}},\ }\href@noop {} {\bibfield
		{journal} {\bibinfo  {journal} {Science}\ }\textbf {\bibinfo {volume}
			{367}},\ \bibinfo {pages} {900} (\bibinfo {year} {2020})}\BibitemShut
	{NoStop}%
	\bibitem [{\citenamefont {Lopes~dos Santos}\ \emph {et~al.}(2007)\citenamefont
		{Lopes~dos Santos}, \citenamefont {Peres},\ and\ \citenamefont
		{Castro~Neto}}]{SantosTBG}%
	\BibitemOpen
	\bibfield  {author} {\bibinfo {author} {\bibfnamefont {J.~M.~B.}\
			\bibnamefont {Lopes~dos Santos}}, \bibinfo {author} {\bibfnamefont
			{N.~M.~R.}\ \bibnamefont {Peres}}, \ and\ \bibinfo {author} {\bibfnamefont
			{A.~H.}\ \bibnamefont {Castro~Neto}},\ }\href@noop {} {\bibfield  {journal}
		{\bibinfo  {journal} {Phys. Rev. Lett.}\ }\textbf {\bibinfo {volume} {99}},\
		\bibinfo {pages} {256802} (\bibinfo {year} {2007})}\BibitemShut {NoStop}%
	\bibitem [{\citenamefont {{Bistritzer}}\ and\ \citenamefont
		{{MacDonald}}(2011)}]{bistritzer2011}%
	\BibitemOpen
	\bibfield  {author} {\bibinfo {author} {\bibfnamefont {R.}~\bibnamefont
			{{Bistritzer}}}\ and\ \bibinfo {author} {\bibfnamefont {A.~H.}\ \bibnamefont
			{{MacDonald}}},\ }\href@noop {} {\bibfield  {journal} {\bibinfo  {journal}
			{Proceedings of the National Academy of Science}\ }\textbf {\bibinfo {volume}
			{108}},\ \bibinfo {pages} {12233} (\bibinfo {year} {2011})}\BibitemShut
	{NoStop}%
	\bibitem [{\citenamefont {Castro~Neto}\ \emph {et~al.}(2009)\citenamefont
		{Castro~Neto}, \citenamefont {Guinea}, \citenamefont {Peres}, \citenamefont
		{Novoselov},\ and\ \citenamefont {Geim}}]{CastroNetoRMP}%
	\BibitemOpen
	\bibfield  {author} {\bibinfo {author} {\bibfnamefont {A.~H.}\ \bibnamefont
			{Castro~Neto}}, \bibinfo {author} {\bibfnamefont {F.}~\bibnamefont {Guinea}},
		\bibinfo {author} {\bibfnamefont {N.~M.~R.}\ \bibnamefont {Peres}}, \bibinfo
		{author} {\bibfnamefont {K.~S.}\ \bibnamefont {Novoselov}}, \ and\ \bibinfo
		{author} {\bibfnamefont {A.~K.}\ \bibnamefont {Geim}},\ }\href {\doibase
		10.1103/RevModPhys.81.109} {\bibfield  {journal} {\bibinfo  {journal} {Rev.
				Mod. Phys.}\ }\textbf {\bibinfo {volume} {81}},\ \bibinfo {pages} {109}
		(\bibinfo {year} {2009})}\BibitemShut {NoStop}%
	\bibitem [{\citenamefont {Kim}\ \emph {et~al.}(2017)\citenamefont {Kim},
		\citenamefont {DaSilva}, \citenamefont {Huang}, \citenamefont {Fallahazad},
		\citenamefont {Larentis}, \citenamefont {Taniguchi}, \citenamefont
		{Watanabe}, \citenamefont {LeRoy}, \citenamefont {MacDonald},\ and\
		\citenamefont {Tutuc}}]{Tutuc_TBG}%
	\BibitemOpen
	\bibfield  {author} {\bibinfo {author} {\bibfnamefont {K.}~\bibnamefont
			{Kim}}, \bibinfo {author} {\bibfnamefont {A.}~\bibnamefont {DaSilva}},
		\bibinfo {author} {\bibfnamefont {S.}~\bibnamefont {Huang}}, \bibinfo
		{author} {\bibfnamefont {B.}~\bibnamefont {Fallahazad}}, \bibinfo {author}
		{\bibfnamefont {S.}~\bibnamefont {Larentis}}, \bibinfo {author}
		{\bibfnamefont {T.}~\bibnamefont {Taniguchi}}, \bibinfo {author}
		{\bibfnamefont {K.}~\bibnamefont {Watanabe}}, \bibinfo {author}
		{\bibfnamefont {B.~J.}\ \bibnamefont {LeRoy}}, \bibinfo {author}
		{\bibfnamefont {A.~H.}\ \bibnamefont {MacDonald}}, \ and\ \bibinfo {author}
		{\bibfnamefont {E.}~\bibnamefont {Tutuc}},\ }\href {\doibase
		10.1073/pnas.1620140114} {\bibfield  {journal} {\bibinfo  {journal}
			{Proceedings of the National Academy of Sciences}\ }\textbf {\bibinfo
			{volume} {114}},\ \bibinfo {pages} {3364} (\bibinfo {year}
		{2017})}\BibitemShut {NoStop}%
	\bibitem [{\citenamefont {{Park}}\ \emph {et~al.}(2020)\citenamefont {{Park}},
		\citenamefont {{Cao}}, \citenamefont {{Watanabe}}, \citenamefont
		{{Taniguchi}},\ and\ \citenamefont {{Jarillo-Herrero}}}]{lasercut}%
	\BibitemOpen
	\bibfield  {author} {\bibinfo {author} {\bibfnamefont {J.~M.}\ \bibnamefont
			{{Park}}}, \bibinfo {author} {\bibfnamefont {Y.}~\bibnamefont {{Cao}}},
		\bibinfo {author} {\bibfnamefont {K.}~\bibnamefont {{Watanabe}}}, \bibinfo
		{author} {\bibfnamefont {T.}~\bibnamefont {{Taniguchi}}}, \ and\ \bibinfo
		{author} {\bibfnamefont {P.}~\bibnamefont {{Jarillo-Herrero}}},\ }\href@noop
	{} {\bibfield  {journal} {\bibinfo  {journal} {arXiv e-prints}\ ,\ \bibinfo
			{eid} {arXiv:2008.12296}} (\bibinfo {year} {2020})},\ \Eprint
	{http://arxiv.org/abs/2008.12296} {arXiv:2008.12296 [cond-mat.mes-hall]}
	\BibitemShut {NoStop}%
	\bibitem [{\citenamefont {Purdie}\ \emph {et~al.}(2018)\citenamefont {Purdie},
		\citenamefont {Pugno}, \citenamefont {Taniguchi}, \citenamefont {Watanabe},
		\citenamefont {Ferrari},\ and\ \citenamefont
		{Lombardo}}]{Hot-transfer_NComm}%
	\BibitemOpen
	\bibfield  {author} {\bibinfo {author} {\bibfnamefont {D.~G.}\ \bibnamefont
			{Purdie}}, \bibinfo {author} {\bibfnamefont {N.~M.}\ \bibnamefont {Pugno}},
		\bibinfo {author} {\bibfnamefont {T.}~\bibnamefont {Taniguchi}}, \bibinfo
		{author} {\bibfnamefont {K.}~\bibnamefont {Watanabe}}, \bibinfo {author}
		{\bibfnamefont {A.~C.}\ \bibnamefont {Ferrari}}, \ and\ \bibinfo {author}
		{\bibfnamefont {A.}~\bibnamefont {Lombardo}},\ }\href {\doibase
		10.1038/s41467-018-07558-3} {\bibfield  {journal} {\bibinfo  {journal}
			{Nature Communications}\ }\textbf {\bibinfo {volume} {9}},\ \bibinfo {pages}
		{5387} (\bibinfo {year} {2018})}\BibitemShut {NoStop}%
	\bibitem [{\citenamefont {Ben~Shalom}\ \emph {et~al.}(2016)\citenamefont
		{Ben~Shalom}, \citenamefont {Zhu}, \citenamefont {Fal'ko}, \citenamefont
		{Mishchenko}, \citenamefont {Kretinin}, \citenamefont {Novoselov},
		\citenamefont {Woods}, \citenamefont {Watanabe}, \citenamefont {Taniguchi},
		\citenamefont {Geim},\ and\ \citenamefont {Prance}}]{MosheNatPhys}%
	\BibitemOpen
	\bibfield  {author} {\bibinfo {author} {\bibfnamefont {M.}~\bibnamefont
			{Ben~Shalom}}, \bibinfo {author} {\bibfnamefont {M.~J.}\ \bibnamefont {Zhu}},
		\bibinfo {author} {\bibfnamefont {V.~I.}\ \bibnamefont {Fal'ko}}, \bibinfo
		{author} {\bibfnamefont {A.}~\bibnamefont {Mishchenko}}, \bibinfo {author}
		{\bibfnamefont {A.~V.}\ \bibnamefont {Kretinin}}, \bibinfo {author}
		{\bibfnamefont {K.~S.}\ \bibnamefont {Novoselov}}, \bibinfo {author}
		{\bibfnamefont {C.~R.}\ \bibnamefont {Woods}}, \bibinfo {author}
		{\bibfnamefont {K.}~\bibnamefont {Watanabe}}, \bibinfo {author}
		{\bibfnamefont {T.}~\bibnamefont {Taniguchi}}, \bibinfo {author}
		{\bibfnamefont {A.~K.}\ \bibnamefont {Geim}}, \ and\ \bibinfo {author}
		{\bibfnamefont {J.~R.}\ \bibnamefont {Prance}},\ }\href {\doibase
		10.1038/nphys3592} {\bibfield  {journal} {\bibinfo  {journal} {Nature
				Physics}\ }\textbf {\bibinfo {volume} {12}},\ \bibinfo {pages} {318}
		(\bibinfo {year} {2016})}\BibitemShut {NoStop}%
	\bibitem [{Sup()}]{Supplementary}%
	\BibitemOpen
	\href@noop {} {\bibinfo  {journal} {See Supplemental Material at
			http://link.aps.org/ supplemental/ for device fabrication details, angle
			determination procedure, connection between the applied gate voltages and
			density imbalance, Interminivalley oscillations under strong dc bias,
			estimating scattering rates ration in SA-TBG.}\ }\BibitemShut {NoStop}%
	\bibitem [{\citenamefont {Kim}\ \emph {et~al.}(2016)\citenamefont {Kim},
		\citenamefont {Herlinger}, \citenamefont {Moon}, \citenamefont {Koshino},
		\citenamefont {Taniguchi}, \citenamefont {Watanabe},\ and\ \citenamefont
		{Smet}}]{SmetTBG}%
	\BibitemOpen
	\bibfield  {journal} {  }\bibfield  {author} {\bibinfo {author} {\bibfnamefont
			{Y.}~\bibnamefont {Kim}}, \bibinfo {author} {\bibfnamefont {P.}~\bibnamefont
			{Herlinger}}, \bibinfo {author} {\bibfnamefont {P.}~\bibnamefont {Moon}},
		\bibinfo {author} {\bibfnamefont {M.}~\bibnamefont {Koshino}}, \bibinfo
		{author} {\bibfnamefont {T.}~\bibnamefont {Taniguchi}}, \bibinfo {author}
		{\bibfnamefont {K.}~\bibnamefont {Watanabe}}, \ and\ \bibinfo {author}
		{\bibfnamefont {J.~H.}\ \bibnamefont {Smet}},\ }\href@noop {} {\bibfield
		{journal} {\bibinfo  {journal} {Nano Letters}\ }\textbf {\bibinfo {volume}
			{16}},\ \bibinfo {pages} {5053} (\bibinfo {year} {2016})}\BibitemShut
	{NoStop}%
	\bibitem [{\citenamefont {Cao}\ \emph {et~al.}(2016)\citenamefont {Cao},
		\citenamefont {Luo}, \citenamefont {Fatemi}, \citenamefont {Fang},
		\citenamefont {Sanchez-Yamagishi}, \citenamefont {Watanabe}, \citenamefont
		{Taniguchi}, \citenamefont {Kaxiras},\ and\ \citenamefont
		{Jarillo-Herrero}}]{YuanPRL}%
	\BibitemOpen
	\bibfield  {author} {\bibinfo {author} {\bibfnamefont {Y.}~\bibnamefont
			{Cao}}, \bibinfo {author} {\bibfnamefont {J.~Y.}\ \bibnamefont {Luo}},
		\bibinfo {author} {\bibfnamefont {V.}~\bibnamefont {Fatemi}}, \bibinfo
		{author} {\bibfnamefont {S.}~\bibnamefont {Fang}}, \bibinfo {author}
		{\bibfnamefont {J.~D.}\ \bibnamefont {Sanchez-Yamagishi}}, \bibinfo {author}
		{\bibfnamefont {K.}~\bibnamefont {Watanabe}}, \bibinfo {author}
		{\bibfnamefont {T.}~\bibnamefont {Taniguchi}}, \bibinfo {author}
		{\bibfnamefont {E.}~\bibnamefont {Kaxiras}}, \ and\ \bibinfo {author}
		{\bibfnamefont {P.}~\bibnamefont {Jarillo-Herrero}},\ }\href {\doibase
		10.1103/PhysRevLett.117.116804} {\bibfield  {journal} {\bibinfo  {journal}
			{Phys. Rev. Lett.}\ }\textbf {\bibinfo {volume} {117}},\ \bibinfo {pages}
		{116804} (\bibinfo {year} {2016})}\BibitemShut {NoStop}%
	\bibitem [{\citenamefont {de~Vries}\ \emph {et~al.}(2020)\citenamefont
		{de~Vries}, \citenamefont {Zhu}, \citenamefont {Portol\'es}, \citenamefont
		{Zheng}, \citenamefont {Masseroni}, \citenamefont {Kurzmann}, \citenamefont
		{Taniguchi}, \citenamefont {Watanabe}, \citenamefont {MacDonald},
		\citenamefont {Ensslin}, \citenamefont {Ihn},\ and\ \citenamefont
		{Rickhaus}}]{Rickhaus_minivalley}%
	\BibitemOpen
	\bibfield  {author} {\bibinfo {author} {\bibfnamefont {F.~K.}\ \bibnamefont
			{de~Vries}}, \bibinfo {author} {\bibfnamefont {J.}~\bibnamefont {Zhu}},
		\bibinfo {author} {\bibfnamefont {E.}~\bibnamefont {Portol\'es}}, \bibinfo
		{author} {\bibfnamefont {G.}~\bibnamefont {Zheng}}, \bibinfo {author}
		{\bibfnamefont {M.}~\bibnamefont {Masseroni}}, \bibinfo {author}
		{\bibfnamefont {A.}~\bibnamefont {Kurzmann}}, \bibinfo {author}
		{\bibfnamefont {T.}~\bibnamefont {Taniguchi}}, \bibinfo {author}
		{\bibfnamefont {K.}~\bibnamefont {Watanabe}}, \bibinfo {author}
		{\bibfnamefont {A.~H.}\ \bibnamefont {MacDonald}}, \bibinfo {author}
		{\bibfnamefont {K.}~\bibnamefont {Ensslin}}, \bibinfo {author} {\bibfnamefont
			{T.}~\bibnamefont {Ihn}}, \ and\ \bibinfo {author} {\bibfnamefont
			{P.}~\bibnamefont {Rickhaus}},\ }\href@noop {} {\bibfield  {journal}
		{\bibinfo  {journal} {Phys. Rev. Lett.}\ }\textbf {\bibinfo {volume} {125}},\
		\bibinfo {pages} {176801} (\bibinfo {year} {2020})}\BibitemShut {NoStop}%
	\bibitem [{\citenamefont {{Berdyugin}}\ \emph {et~al.}(2020)\citenamefont
		{{Berdyugin}}, \citenamefont {{Tsim}}, \citenamefont {{Kumaravadivel}},
		\citenamefont {{Xu}}, \citenamefont {{Ceferino}}, \citenamefont {{Knothe}},
		\citenamefont {{Kumar}}, \citenamefont {{Taniguchi}}, \citenamefont
		{{Watanabe}}, \citenamefont {{Geim}}, \citenamefont {{Grigorieva}},\ and\
		\citenamefont {{Fal'ko}}}]{berdyugin2020}%
	\BibitemOpen
	\bibfield  {author} {\bibinfo {author} {\bibfnamefont {A.~I.}\ \bibnamefont
			{{Berdyugin}}}, \bibinfo {author} {\bibfnamefont {B.}~\bibnamefont {{Tsim}}},
		\bibinfo {author} {\bibfnamefont {P.}~\bibnamefont {{Kumaravadivel}}},
		\bibinfo {author} {\bibfnamefont {S.~G.}\ \bibnamefont {{Xu}}}, \bibinfo
		{author} {\bibfnamefont {A.}~\bibnamefont {{Ceferino}}}, \bibinfo {author}
		{\bibfnamefont {A.}~\bibnamefont {{Knothe}}}, \bibinfo {author}
		{\bibfnamefont {R.~K.}\ \bibnamefont {{Kumar}}}, \bibinfo {author}
		{\bibfnamefont {T.}~\bibnamefont {{Taniguchi}}}, \bibinfo {author}
		{\bibfnamefont {K.}~\bibnamefont {{Watanabe}}}, \bibinfo {author}
		{\bibfnamefont {A.~K.}\ \bibnamefont {{Geim}}}, \bibinfo {author}
		{\bibfnamefont {I.~V.}\ \bibnamefont {{Grigorieva}}}, \ and\ \bibinfo
		{author} {\bibfnamefont {V.~I.}\ \bibnamefont {{Fal'ko}}},\ }\href@noop {}
	{\bibfield  {journal} {\bibinfo  {journal} {Science Advances}\ }\textbf
		{\bibinfo {volume} {6}},\ \bibinfo {pages} {eaay7838} (\bibinfo {year}
		{2020})}\BibitemShut {NoStop}%
	\bibitem [{\citenamefont {Slizovskiy}\ \emph {et~al.}(2019)\citenamefont
		{Slizovskiy}, \citenamefont {Garcia-Ruiz}, \citenamefont {Drummond},\ and\
		\citenamefont {Falko}}]{VolodyaScreening}%
	\BibitemOpen
	\bibfield  {author} {\bibinfo {author} {\bibfnamefont {S.}~\bibnamefont
			{Slizovskiy}}, \bibinfo {author} {\bibfnamefont {A.}~\bibnamefont
			{Garcia-Ruiz}}, \bibinfo {author} {\bibfnamefont {N.}~\bibnamefont
			{Drummond}}, \ and\ \bibinfo {author} {\bibfnamefont {V.~I.}\ \bibnamefont
			{Falko}},\ }\href@noop {} {\enquote {\bibinfo {title} {Dielectric
				susceptibility of graphene describing its out-of-plane polarizability},}\ }
	(\bibinfo {year} {2019}),\ \Eprint {http://arxiv.org/abs/1912.10067}
	{arXiv:1912.10067 [cond-mat.mes-hall]} \BibitemShut {NoStop}%
	\bibitem [{\citenamefont {Sanchez-Yamagishi}\ \emph {et~al.}(2012)\citenamefont
		{Sanchez-Yamagishi}, \citenamefont {Taychatanapat}, \citenamefont {Watanabe},
		\citenamefont {Taniguchi}, \citenamefont {Yacoby},\ and\ \citenamefont
		{Jarillo-Herrero}}]{SanchezPRL}%
	\BibitemOpen
	\bibfield  {author} {\bibinfo {author} {\bibfnamefont {J.~D.}\ \bibnamefont
			{Sanchez-Yamagishi}}, \bibinfo {author} {\bibfnamefont {T.}~\bibnamefont
			{Taychatanapat}}, \bibinfo {author} {\bibfnamefont {K.}~\bibnamefont
			{Watanabe}}, \bibinfo {author} {\bibfnamefont {T.}~\bibnamefont {Taniguchi}},
		\bibinfo {author} {\bibfnamefont {A.}~\bibnamefont {Yacoby}}, \ and\ \bibinfo
		{author} {\bibfnamefont {P.}~\bibnamefont {Jarillo-Herrero}},\ }\href
	{\doibase 10.1103/PhysRevLett.108.076601} {\bibfield  {journal} {\bibinfo
			{journal} {Phys. Rev. Lett.}\ }\textbf {\bibinfo {volume} {108}},\ \bibinfo
		{pages} {076601} (\bibinfo {year} {2012})}\BibitemShut {NoStop}%
	\bibitem [{\citenamefont {Fallahazad}\ \emph {et~al.}(2012)\citenamefont
		{Fallahazad}, \citenamefont {Hao}, \citenamefont {Lee}, \citenamefont {Kim},
		\citenamefont {Ruoff},\ and\ \citenamefont {Tutuc}}]{FallahazadPRB}%
	\BibitemOpen
	\bibfield  {author} {\bibinfo {author} {\bibfnamefont {B.}~\bibnamefont
			{Fallahazad}}, \bibinfo {author} {\bibfnamefont {Y.}~\bibnamefont {Hao}},
		\bibinfo {author} {\bibfnamefont {K.}~\bibnamefont {Lee}}, \bibinfo {author}
		{\bibfnamefont {S.}~\bibnamefont {Kim}}, \bibinfo {author} {\bibfnamefont
			{R.~S.}\ \bibnamefont {Ruoff}}, \ and\ \bibinfo {author} {\bibfnamefont
			{E.}~\bibnamefont {Tutuc}},\ }\href {\doibase 10.1103/PhysRevB.85.201408}
	{\bibfield  {journal} {\bibinfo  {journal} {Phys. Rev. B}\ }\textbf {\bibinfo
			{volume} {85}},\ \bibinfo {pages} {201408} (\bibinfo {year}
		{2012})}\BibitemShut {NoStop}%
	\bibitem [{\citenamefont {Schmidt}\ \emph {et~al.}(2008)\citenamefont
		{Schmidt}, \citenamefont {Lüdtke}, \citenamefont {Barthold}, \citenamefont
		{McCann}, \citenamefont {Fal’ko},\ and\ \citenamefont
		{Haug}}]{VolodyaBeating}%
	\BibitemOpen
	\bibfield  {author} {\bibinfo {author} {\bibfnamefont {H.}~\bibnamefont
			{Schmidt}}, \bibinfo {author} {\bibfnamefont {T.}~\bibnamefont {Lüdtke}},
		\bibinfo {author} {\bibfnamefont {P.}~\bibnamefont {Barthold}}, \bibinfo
		{author} {\bibfnamefont {E.}~\bibnamefont {McCann}}, \bibinfo {author}
		{\bibfnamefont {V.~I.}\ \bibnamefont {Fal’ko}}, \ and\ \bibinfo {author}
		{\bibfnamefont {R.~J.}\ \bibnamefont {Haug}},\ }\href@noop {} {\bibfield
		{journal} {\bibinfo  {journal} {Applied Physics Letters}\ }\textbf {\bibinfo
			{volume} {93}},\ \bibinfo {pages} {172108} (\bibinfo {year}
		{2008})}\BibitemShut {NoStop}%
	\bibitem [{\citenamefont {Chung}\ \emph
		{et~al.}(2018{\natexlab{a}})\citenamefont {Chung}, \citenamefont {Xu},\ and\
		\citenamefont {Chen}}]{Chung2018}%
	\BibitemOpen
	\bibfield  {author} {\bibinfo {author} {\bibfnamefont {T.-F.}\ \bibnamefont
			{Chung}}, \bibinfo {author} {\bibfnamefont {Y.}~\bibnamefont {Xu}}, \ and\
		\bibinfo {author} {\bibfnamefont {Y.~P.}\ \bibnamefont {Chen}},\ }\href
	{\doibase 10.1103/PhysRevB.98.035425} {\bibfield  {journal} {\bibinfo
			{journal} {Phys. Rev. B}\ }\textbf {\bibinfo {volume} {98}},\ \bibinfo
		{pages} {035425} (\bibinfo {year} {2018}{\natexlab{a}})}\BibitemShut
	{NoStop}%
	\bibitem [{\citenamefont {Coleridge}(1990)}]{Coleridge}%
	\BibitemOpen
	\bibfield  {author} {\bibinfo {author} {\bibfnamefont {P.~T.}\ \bibnamefont
			{Coleridge}},\ }\href@noop {} {\bibfield  {journal} {\bibinfo  {journal}
			{Semiconductor Science and Technology}\ }\textbf {\bibinfo {volume} {5}},\
		\bibinfo {pages} {961} (\bibinfo {year} {1990})}\BibitemShut {NoStop}%
	\bibitem [{\citenamefont {Polyanovsky}(1988)}]{Polyanovsky}%
	\BibitemOpen
	\bibfield  {author} {\bibinfo {author} {\bibfnamefont {V.}~\bibnamefont
			{Polyanovsky}},\ }\href@noop {} {\bibfield  {journal} {\bibinfo  {journal}
			{Fiz. Tekh. Poluprovodn.}\ }\textbf {\bibinfo {volume} {12}},\ \bibinfo
		{pages} {2230} (\bibinfo {year} {1988})}\BibitemShut {NoStop}%
	\bibitem [{\citenamefont {Leadley}\ \emph {et~al.}(1992)\citenamefont
		{Leadley}, \citenamefont {Fletcher}, \citenamefont {Nicholas}, \citenamefont
		{Tao}, \citenamefont {Foxon},\ and\ \citenamefont {Harris}}]{MISO_old}%
	\BibitemOpen
	\bibfield  {author} {\bibinfo {author} {\bibfnamefont {D.~R.}\ \bibnamefont
			{Leadley}}, \bibinfo {author} {\bibfnamefont {R.}~\bibnamefont {Fletcher}},
		\bibinfo {author} {\bibfnamefont {R.~J.}\ \bibnamefont {Nicholas}}, \bibinfo
		{author} {\bibfnamefont {F.}~\bibnamefont {Tao}}, \bibinfo {author}
		{\bibfnamefont {C.~T.}\ \bibnamefont {Foxon}}, \ and\ \bibinfo {author}
		{\bibfnamefont {J.~J.}\ \bibnamefont {Harris}},\ }\href@noop {} {\bibfield
		{journal} {\bibinfo  {journal} {Phys. Rev. B}\ }\textbf {\bibinfo {volume}
			{46}},\ \bibinfo {pages} {12439} (\bibinfo {year} {1992})}\BibitemShut
	{NoStop}%
	\bibitem [{\citenamefont {Raichev}(2008)}]{Raichev2008}%
	\BibitemOpen
	\bibfield  {author} {\bibinfo {author} {\bibfnamefont {O.~E.}\ \bibnamefont
			{Raichev}},\ }\href {\doibase 10.1103/PhysRevB.78.125304} {\bibfield
		{journal} {\bibinfo  {journal} {Phys. Rev. B}\ }\textbf {\bibinfo {volume}
			{78}},\ \bibinfo {pages} {125304} (\bibinfo {year} {2008})}\BibitemShut
	{NoStop}%
	\bibitem [{\citenamefont {Mamani}\ \emph
		{et~al.}(2009{\natexlab{a}})\citenamefont {Mamani}, \citenamefont {Gusev},
		\citenamefont {da~Silva}, \citenamefont {Raichev}, \citenamefont {Quivy},\
		and\ \citenamefont {Bakarov}}]{RaichevMISO}%
	\BibitemOpen
	\bibfield  {author} {\bibinfo {author} {\bibfnamefont {N.~C.}\ \bibnamefont
			{Mamani}}, \bibinfo {author} {\bibfnamefont {G.~M.}\ \bibnamefont {Gusev}},
		\bibinfo {author} {\bibfnamefont {E.~C.~F.}\ \bibnamefont {da~Silva}},
		\bibinfo {author} {\bibfnamefont {O.~E.}\ \bibnamefont {Raichev}}, \bibinfo
		{author} {\bibfnamefont {A.~A.}\ \bibnamefont {Quivy}}, \ and\ \bibinfo
		{author} {\bibfnamefont {A.~K.}\ \bibnamefont {Bakarov}},\ }\href@noop {}
	{\bibfield  {journal} {\bibinfo  {journal} {Phys. Rev. B}\ }\textbf {\bibinfo
			{volume} {80}},\ \bibinfo {pages} {085304} (\bibinfo {year}
		{2009}{\natexlab{a}})}\BibitemShut {NoStop}%
	\bibitem [{\citenamefont {{Kartsovnik}}\ \emph {et~al.}(2002)\citenamefont
		{{Kartsovnik}}, \citenamefont {{Grigoriev}}, \citenamefont {{Biberacher}},
		\citenamefont {{Kushch}},\ and\ \citenamefont {{Wyder}}}]{kartsovnik2002}%
	\BibitemOpen
	\bibfield  {author} {\bibinfo {author} {\bibfnamefont {M.~V.}\ \bibnamefont
			{{Kartsovnik}}}, \bibinfo {author} {\bibfnamefont {P.~D.}\ \bibnamefont
			{{Grigoriev}}}, \bibinfo {author} {\bibfnamefont {W.}~\bibnamefont
			{{Biberacher}}}, \bibinfo {author} {\bibfnamefont {N.~D.}\ \bibnamefont
			{{Kushch}}}, \ and\ \bibinfo {author} {\bibfnamefont {P.}~\bibnamefont
			{{Wyder}}},\ }\href@noop {} {\bibfield  {journal} {\bibinfo  {journal}
			{\prl}\ }\textbf {\bibinfo {volume} {89}},\ \bibinfo {eid} {126802} (\bibinfo
		{year} {2002})}\BibitemShut {NoStop}%
	\bibitem [{\citenamefont {Minkov}\ \emph {et~al.}(2020)\citenamefont {Minkov},
		\citenamefont {Rut}, \citenamefont {Sherstobitov}, \citenamefont {Dvoretski},
		\citenamefont {Mikhailov}, \citenamefont {Solov'ev}, \citenamefont {Chernov},
		\citenamefont {Ivanov},\ and\ \citenamefont {Germanenko}}]{Minkov}%
	\BibitemOpen
	\bibfield  {author} {\bibinfo {author} {\bibfnamefont {G.~M.}\ \bibnamefont
			{Minkov}}, \bibinfo {author} {\bibfnamefont {O.~E.}\ \bibnamefont {Rut}},
		\bibinfo {author} {\bibfnamefont {A.~A.}\ \bibnamefont {Sherstobitov}},
		\bibinfo {author} {\bibfnamefont {S.~A.}\ \bibnamefont {Dvoretski}}, \bibinfo
		{author} {\bibfnamefont {N.~N.}\ \bibnamefont {Mikhailov}}, \bibinfo {author}
		{\bibfnamefont {V.~A.}\ \bibnamefont {Solov'ev}}, \bibinfo {author}
		{\bibfnamefont {M.~Y.}\ \bibnamefont {Chernov}}, \bibinfo {author}
		{\bibfnamefont {S.~V.}\ \bibnamefont {Ivanov}}, \ and\ \bibinfo {author}
		{\bibfnamefont {A.~V.}\ \bibnamefont {Germanenko}},\ }\href@noop {}
	{\bibfield  {journal} {\bibinfo  {journal} {Phys. Rev. B}\ }\textbf {\bibinfo
			{volume} {101}},\ \bibinfo {pages} {245303} (\bibinfo {year}
		{2020})}\BibitemShut {NoStop}%
	\bibitem [{\citenamefont {Abedi}\ \emph {et~al.}(2021)\citenamefont {Abedi},
		\citenamefont {Vitkalov}, \citenamefont {Bykov},\ and\ \citenamefont
		{Bakarov}}]{Vitkalov_Bparallel}%
	\BibitemOpen
	\bibfield  {author} {\bibinfo {author} {\bibfnamefont {S.}~\bibnamefont
			{Abedi}}, \bibinfo {author} {\bibfnamefont {S.~A.}\ \bibnamefont {Vitkalov}},
		\bibinfo {author} {\bibfnamefont {A.~A.}\ \bibnamefont {Bykov}}, \ and\
		\bibinfo {author} {\bibfnamefont {A.~K.}\ \bibnamefont {Bakarov}},\
	}\href@noop {} {\enquote {\bibinfo {title} {Temperature damping of
				magneto-intersubband resistance oscillations in magnetically entangled
				subbands},}\ } (\bibinfo {year} {2021}),\ \Eprint
	{http://arxiv.org/abs/2105.12263} {arXiv:2105.12263 [cond-mat.mes-hall]}
	\BibitemShut {NoStop}%
	\bibitem [{\citenamefont {Dmitriev}\ \emph {et~al.}(2012)\citenamefont
		{Dmitriev}, \citenamefont {Mirlin}, \citenamefont {Polyakov},\ and\
		\citenamefont {Zudov}}]{RevModPhysVanya}%
	\BibitemOpen
	\bibfield  {author} {\bibinfo {author} {\bibfnamefont {I.~A.}\ \bibnamefont
			{Dmitriev}}, \bibinfo {author} {\bibfnamefont {A.~D.}\ \bibnamefont
			{Mirlin}}, \bibinfo {author} {\bibfnamefont {D.~G.}\ \bibnamefont
			{Polyakov}}, \ and\ \bibinfo {author} {\bibfnamefont {M.~A.}\ \bibnamefont
			{Zudov}},\ }\href {\doibase 10.1103/RevModPhys.84.1709} {\bibfield  {journal}
		{\bibinfo  {journal} {Rev. Mod. Phys.}\ }\textbf {\bibinfo {volume} {84}},\
		\bibinfo {pages} {1709} (\bibinfo {year} {2012})}\BibitemShut {NoStop}%
	\bibitem [{\citenamefont {M{\"{o}}nch}\ \emph {et~al.}(2020)\citenamefont
		{M{\"{o}}nch}, \citenamefont {Bandurin}, \citenamefont {Dmitriev},
		\citenamefont {Phinney}, \citenamefont {Yahniuk}, \citenamefont {Taniguchi},
		\citenamefont {Watanabe}, \citenamefont {Jarillo-Herrero},\ and\
		\citenamefont {Ganichev}}]{TIMO}%
	\BibitemOpen
	\bibfield  {author} {\bibinfo {author} {\bibfnamefont {E.}~\bibnamefont
			{M{\"{o}}nch}}, \bibinfo {author} {\bibfnamefont {D.~A.}\ \bibnamefont
			{Bandurin}}, \bibinfo {author} {\bibfnamefont {I.~A.}\ \bibnamefont
			{Dmitriev}}, \bibinfo {author} {\bibfnamefont {I.~Y.}\ \bibnamefont
			{Phinney}}, \bibinfo {author} {\bibfnamefont {I.}~\bibnamefont {Yahniuk}},
		\bibinfo {author} {\bibfnamefont {T.}~\bibnamefont {Taniguchi}}, \bibinfo
		{author} {\bibfnamefont {K.}~\bibnamefont {Watanabe}}, \bibinfo {author}
		{\bibfnamefont {P.}~\bibnamefont {Jarillo-Herrero}}, \ and\ \bibinfo {author}
		{\bibfnamefont {S.~D.}\ \bibnamefont {Ganichev}},\ }\href {\doibase
		10.1021/acs.nanolett.0c01918} {\bibfield  {journal} {\bibinfo  {journal}
			{Nano Letters}\ }\textbf {\bibinfo {volume} {20}},\ \bibinfo {pages} {5943}
		(\bibinfo {year} {2020})}\BibitemShut {NoStop}%
	\bibitem [{\citenamefont {Bykov}(2008)}]{Bykov2008}%
	\BibitemOpen
	\bibfield  {author} {\bibinfo {author} {\bibfnamefont {A.~A.}\ \bibnamefont
			{Bykov}},\ }\href {\doibase 10.1134/S0021364008130146} {\bibfield  {journal}
		{\bibinfo  {journal} {JETP Letters}\ }\textbf {\bibinfo {volume} {88}},\
		\bibinfo {pages} {64} (\bibinfo {year} {2008})}\BibitemShut {NoStop}%
	\bibitem [{\citenamefont {Mamani}\ \emph
		{et~al.}(2009{\natexlab{b}})\citenamefont {Mamani}, \citenamefont {Gusev},
		\citenamefont {Raichev}, \citenamefont {Lamas},\ and\ \citenamefont
		{Bakarov}}]{HIRO_MISO}%
	\BibitemOpen
	\bibfield  {author} {\bibinfo {author} {\bibfnamefont {N.~C.}\ \bibnamefont
			{Mamani}}, \bibinfo {author} {\bibfnamefont {G.~M.}\ \bibnamefont {Gusev}},
		\bibinfo {author} {\bibfnamefont {O.~E.}\ \bibnamefont {Raichev}}, \bibinfo
		{author} {\bibfnamefont {T.~E.}\ \bibnamefont {Lamas}}, \ and\ \bibinfo
		{author} {\bibfnamefont {A.~K.}\ \bibnamefont {Bakarov}},\ }\href@noop {}
	{\bibfield  {journal} {\bibinfo  {journal} {Phys. Rev. B}\ }\textbf {\bibinfo
			{volume} {80}},\ \bibinfo {pages} {075308} (\bibinfo {year}
		{2009}{\natexlab{b}})}\BibitemShut {NoStop}%
	\bibitem [{\citenamefont {Wiedmann}\ \emph {et~al.}(2011)\citenamefont
		{Wiedmann}, \citenamefont {Gusev}, \citenamefont {Raichev}, \citenamefont
		{Bakarov},\ and\ \citenamefont {Portal}}]{Wiedmann2011}%
	\BibitemOpen
	\bibfield  {author} {\bibinfo {author} {\bibfnamefont {S.}~\bibnamefont
			{Wiedmann}}, \bibinfo {author} {\bibfnamefont {G.~M.}\ \bibnamefont {Gusev}},
		\bibinfo {author} {\bibfnamefont {O.~E.}\ \bibnamefont {Raichev}}, \bibinfo
		{author} {\bibfnamefont {A.~K.}\ \bibnamefont {Bakarov}}, \ and\ \bibinfo
		{author} {\bibfnamefont {J.~C.}\ \bibnamefont {Portal}},\ }\href {\doibase
		10.1103/PhysRevB.84.165303} {\bibfield  {journal} {\bibinfo  {journal} {Phys.
				Rev. B}\ }\textbf {\bibinfo {volume} {84}},\ \bibinfo {pages} {165303}
		(\bibinfo {year} {2011})}\BibitemShut {NoStop}%
	\bibitem [{\citenamefont {Drichko}\ \emph {et~al.}(2020)\citenamefont
		{Drichko}, \citenamefont {Smirnov}, \citenamefont {Bakarov}, \citenamefont
		{Bykov}, \citenamefont {Dmitriev},\ and\ \citenamefont
		{Galperin}}]{Drichko2020}%
	\BibitemOpen
	\bibfield  {author} {\bibinfo {author} {\bibfnamefont {I.~L.}\ \bibnamefont
			{Drichko}}, \bibinfo {author} {\bibfnamefont {I.~Y.}\ \bibnamefont
			{Smirnov}}, \bibinfo {author} {\bibfnamefont {A.~K.}\ \bibnamefont
			{Bakarov}}, \bibinfo {author} {\bibfnamefont {A.~A.}\ \bibnamefont {Bykov}},
		\bibinfo {author} {\bibfnamefont {A.~A.}\ \bibnamefont {Dmitriev}}, \ and\
		\bibinfo {author} {\bibfnamefont {Y.~M.}\ \bibnamefont {Galperin}},\ }\href
	{\doibase 10.1134/S0021364020130068} {\bibfield  {journal} {\bibinfo
			{journal} {JETP Letters}\ }\textbf {\bibinfo {volume} {112}},\ \bibinfo
		{pages} {45} (\bibinfo {year} {2020})}\BibitemShut {NoStop}%
	\bibitem [{\citenamefont {Yang}\ \emph {et~al.}(2002)\citenamefont {Yang},
		\citenamefont {Zhang}, \citenamefont {Du}, \citenamefont {Simmons},\ and\
		\citenamefont {Reno}}]{Reno}%
	\BibitemOpen
	\bibfield  {author} {\bibinfo {author} {\bibfnamefont {C.~L.}\ \bibnamefont
			{Yang}}, \bibinfo {author} {\bibfnamefont {J.}~\bibnamefont {Zhang}},
		\bibinfo {author} {\bibfnamefont {R.~R.}\ \bibnamefont {Du}}, \bibinfo
		{author} {\bibfnamefont {J.~A.}\ \bibnamefont {Simmons}}, \ and\ \bibinfo
		{author} {\bibfnamefont {J.~L.}\ \bibnamefont {Reno}},\ }\href {\doibase
		10.1103/PhysRevLett.89.076801} {\bibfield  {journal} {\bibinfo  {journal}
			{Phys. Rev. Lett.}\ }\textbf {\bibinfo {volume} {89}},\ \bibinfo {pages}
		{076801} (\bibinfo {year} {2002})}\BibitemShut {NoStop}%
	\bibitem [{\citenamefont {Vavilov}\ \emph {et~al.}(2007)\citenamefont
		{Vavilov}, \citenamefont {Aleiner},\ and\ \citenamefont {Glazman}}]{Glazman}%
	\BibitemOpen
	\bibfield  {author} {\bibinfo {author} {\bibfnamefont {M.~G.}\ \bibnamefont
			{Vavilov}}, \bibinfo {author} {\bibfnamefont {I.~L.}\ \bibnamefont
			{Aleiner}}, \ and\ \bibinfo {author} {\bibfnamefont {L.~I.}\ \bibnamefont
			{Glazman}},\ }\href {\doibase 10.1103/PhysRevB.76.115331} {\bibfield
		{journal} {\bibinfo  {journal} {Phys. Rev. B}\ }\textbf {\bibinfo {volume}
			{76}},\ \bibinfo {pages} {115331} (\bibinfo {year} {2007})}\BibitemShut
	{NoStop}%
	\bibitem [{\citenamefont {Raichev}\ and\ \citenamefont
		{Zudov}(2020)}]{ZudovGraphene}%
	\BibitemOpen
	\bibfield  {author} {\bibinfo {author} {\bibfnamefont {O.~E.}\ \bibnamefont
			{Raichev}}\ and\ \bibinfo {author} {\bibfnamefont {M.~A.}\ \bibnamefont
			{Zudov}},\ }\href@noop {} {\bibfield  {journal} {\bibinfo  {journal} {Phys.
				Rev. Research}\ }\textbf {\bibinfo {volume} {2}},\ \bibinfo {pages} {022011}
		(\bibinfo {year} {2020})}\BibitemShut {NoStop}%
	\bibitem [{\citenamefont {Goran}\ \emph {et~al.}(2009)\citenamefont {Goran},
		\citenamefont {Bykov}, \citenamefont {Toropov},\ and\ \citenamefont
		{Vitkalov}}]{Vitkalov_ee}%
	\BibitemOpen
	\bibfield  {author} {\bibinfo {author} {\bibfnamefont {A.~V.}\ \bibnamefont
			{Goran}}, \bibinfo {author} {\bibfnamefont {A.~A.}\ \bibnamefont {Bykov}},
		\bibinfo {author} {\bibfnamefont {A.~I.}\ \bibnamefont {Toropov}}, \ and\
		\bibinfo {author} {\bibfnamefont {S.~A.}\ \bibnamefont {Vitkalov}},\ }\href
	{\doibase 10.1103/PhysRevB.80.193305} {\bibfield  {journal} {\bibinfo
			{journal} {Phys. Rev. B}\ }\textbf {\bibinfo {volume} {80}},\ \bibinfo
		{pages} {193305} (\bibinfo {year} {2009})}\BibitemShut {NoStop}%
	\bibitem [{\citenamefont {Polini}\ and\ \citenamefont
		{Vignale}(2014)}]{polini2014quasiparticle}%
	\BibitemOpen
	\bibfield  {author} {\bibinfo {author} {\bibfnamefont {M.}~\bibnamefont
			{Polini}}\ and\ \bibinfo {author} {\bibfnamefont {G.}~\bibnamefont
			{Vignale}},\ }\href@noop {} {} (\bibinfo {year} {2014}),\ \Eprint
	{http://arxiv.org/abs/1404.5728} {arXiv:1404.5728 [cond-mat.mes-hall]}
	\BibitemShut {NoStop}%
	\bibitem [{\citenamefont {{Krishna Kumar}}\ \emph {et~al.}(2017)\citenamefont
		{{Krishna Kumar}}, \citenamefont {{Bandurin}}, \citenamefont {{Pellegrino}},
		\citenamefont {{Cao}}, \citenamefont {{Principi}}, \citenamefont {{Guo}},
		\citenamefont {{Auton}}, \citenamefont {{Ben Shalom}}, \citenamefont
		{{Ponomarenko}}, \citenamefont {{Falkovich}}, \citenamefont {{Watanabe}},
		\citenamefont {{Taniguchi}}, \citenamefont {{Grigorieva}}, \citenamefont
		{{Levitov}}, \citenamefont {{Polini}},\ and\ \citenamefont
		{{Geim}}}]{kumar2017superball}%
	\BibitemOpen
	\bibfield  {author} {\bibinfo {author} {\bibfnamefont {R.}~\bibnamefont
			{{Krishna Kumar}}}, \bibinfo {author} {\bibfnamefont {D.~A.}\ \bibnamefont
			{{Bandurin}}}, \bibinfo {author} {\bibfnamefont {F.~M.~D.}\ \bibnamefont
			{{Pellegrino}}}, \bibinfo {author} {\bibfnamefont {Y.}~\bibnamefont {{Cao}}},
		\bibinfo {author} {\bibfnamefont {A.}~\bibnamefont {{Principi}}}, \bibinfo
		{author} {\bibfnamefont {H.}~\bibnamefont {{Guo}}}, \bibinfo {author}
		{\bibfnamefont {G.~H.}\ \bibnamefont {{Auton}}}, \bibinfo {author}
		{\bibfnamefont {M.}~\bibnamefont {{Ben Shalom}}}, \bibinfo {author}
		{\bibfnamefont {L.~A.}\ \bibnamefont {{Ponomarenko}}}, \bibinfo {author}
		{\bibfnamefont {G.}~\bibnamefont {{Falkovich}}}, \bibinfo {author}
		{\bibfnamefont {K.}~\bibnamefont {{Watanabe}}}, \bibinfo {author}
		{\bibfnamefont {T.}~\bibnamefont {{Taniguchi}}}, \bibinfo {author}
		{\bibfnamefont {I.~V.}\ \bibnamefont {{Grigorieva}}}, \bibinfo {author}
		{\bibfnamefont {L.~S.}\ \bibnamefont {{Levitov}}}, \bibinfo {author}
		{\bibfnamefont {M.}~\bibnamefont {{Polini}}}, \ and\ \bibinfo {author}
		{\bibfnamefont {A.~K.}\ \bibnamefont {{Geim}}},\ }\href@noop {} {\bibfield
		{journal} {\bibinfo  {journal} {Nature Physics}\ }\textbf {\bibinfo {volume}
			{13}},\ \bibinfo {pages} {1182} (\bibinfo {year} {2017})}\BibitemShut
	{NoStop}%
	\bibitem [{\citenamefont {Uri}\ \emph {et~al.}(2020)\citenamefont {Uri},
		\citenamefont {Grover}, \citenamefont {Cao}, \citenamefont {Crosse},
		\citenamefont {Bagani}, \citenamefont {Rodan-Legrain}, \citenamefont
		{Myasoedov}, \citenamefont {Watanabe}, \citenamefont {Taniguchi},
		\citenamefont {Moon}, \citenamefont {Koshino}, \citenamefont
		{Jarillo-Herrero},\ and\ \citenamefont {Zeldov}}]{Twistdisorder}%
	\BibitemOpen
	\bibfield  {author} {\bibinfo {author} {\bibfnamefont {A.}~\bibnamefont
			{Uri}}, \bibinfo {author} {\bibfnamefont {S.}~\bibnamefont {Grover}},
		\bibinfo {author} {\bibfnamefont {Y.}~\bibnamefont {Cao}}, \bibinfo {author}
		{\bibfnamefont {J.}~\bibnamefont {Crosse}}, \bibinfo {author} {\bibfnamefont
			{K.}~\bibnamefont {Bagani}}, \bibinfo {author} {\bibfnamefont
			{D.}~\bibnamefont {Rodan-Legrain}}, \bibinfo {author} {\bibfnamefont
			{Y.}~\bibnamefont {Myasoedov}}, \bibinfo {author} {\bibfnamefont
			{K.}~\bibnamefont {Watanabe}}, \bibinfo {author} {\bibfnamefont
			{T.}~\bibnamefont {Taniguchi}}, \bibinfo {author} {\bibfnamefont
			{P.}~\bibnamefont {Moon}}, \bibinfo {author} {\bibfnamefont {M.}~\bibnamefont
			{Koshino}}, \bibinfo {author} {\bibfnamefont {P.}~\bibnamefont
			{Jarillo-Herrero}}, \ and\ \bibinfo {author} {\bibfnamefont {E.}~\bibnamefont
			{Zeldov}},\ }\href@noop {} {\bibfield  {journal} {\bibinfo  {journal}
			{Nature}\ }\textbf {\bibinfo {volume} {581}},\ \bibinfo {pages} {47}
		(\bibinfo {year} {2020})}\BibitemShut {NoStop}%
	\bibitem [{\citenamefont {Raichev}(2010)}]{RaichevMISOphonons}%
	\BibitemOpen
	\bibfield  {author} {\bibinfo {author} {\bibfnamefont {O.~E.}\ \bibnamefont
			{Raichev}},\ }\href@noop {} {\bibfield  {journal} {\bibinfo  {journal} {Phys.
				Rev. B}\ }\textbf {\bibinfo {volume} {81}},\ \bibinfo {pages} {195301}
		(\bibinfo {year} {2010})}\BibitemShut {NoStop}%
	\bibitem [{\citenamefont {Kumaravadivel}\ \emph {et~al.}(2019)\citenamefont
		{Kumaravadivel}, \citenamefont {Greenaway}, \citenamefont {Perello},
		\citenamefont {Berdyugin}, \citenamefont {Birkbeck}, \citenamefont {Wengraf},
		\citenamefont {Liu}, \citenamefont {Edgar}, \citenamefont {Geim},
		\citenamefont {Eaves},\ and\ \citenamefont {Krishna~Kumar}}]{RoshMP}%
	\BibitemOpen
	\bibfield  {author} {\bibinfo {author} {\bibfnamefont {P.}~\bibnamefont
			{Kumaravadivel}}, \bibinfo {author} {\bibfnamefont {M.~T.}\ \bibnamefont
			{Greenaway}}, \bibinfo {author} {\bibfnamefont {D.}~\bibnamefont {Perello}},
		\bibinfo {author} {\bibfnamefont {A.}~\bibnamefont {Berdyugin}}, \bibinfo
		{author} {\bibfnamefont {J.}~\bibnamefont {Birkbeck}}, \bibinfo {author}
		{\bibfnamefont {J.}~\bibnamefont {Wengraf}}, \bibinfo {author} {\bibfnamefont
			{S.}~\bibnamefont {Liu}}, \bibinfo {author} {\bibfnamefont {J.~H.}\
			\bibnamefont {Edgar}}, \bibinfo {author} {\bibfnamefont {A.~K.}\ \bibnamefont
			{Geim}}, \bibinfo {author} {\bibfnamefont {L.}~\bibnamefont {Eaves}}, \ and\
		\bibinfo {author} {\bibfnamefont {R.}~\bibnamefont {Krishna~Kumar}},\ }\href
	{\doibase 10.1038/s41467-019-11379-3} {\bibfield  {journal} {\bibinfo
			{journal} {Nature Communications}\ }\textbf {\bibinfo {volume} {10}},\
		\bibinfo {pages} {3334} (\bibinfo {year} {2019})}\BibitemShut {NoStop}%
	\bibitem [{\citenamefont {Greenaway}\ \emph {et~al.}(2019)\citenamefont
		{Greenaway}, \citenamefont {Krishna~Kumar}, \citenamefont {Kumaravadivel},
		\citenamefont {Geim},\ and\ \citenamefont {Eaves}}]{MarkMP}%
	\BibitemOpen
	\bibfield  {author} {\bibinfo {author} {\bibfnamefont {M.~T.}\ \bibnamefont
			{Greenaway}}, \bibinfo {author} {\bibfnamefont {R.}~\bibnamefont
			{Krishna~Kumar}}, \bibinfo {author} {\bibfnamefont {P.}~\bibnamefont
			{Kumaravadivel}}, \bibinfo {author} {\bibfnamefont {A.~K.}\ \bibnamefont
			{Geim}}, \ and\ \bibinfo {author} {\bibfnamefont {L.}~\bibnamefont {Eaves}},\
	}\href {\doibase 10.1103/PhysRevB.100.155120} {\bibfield  {journal} {\bibinfo
			{journal} {Phys. Rev. B}\ }\textbf {\bibinfo {volume} {100}},\ \bibinfo
		{pages} {155120} (\bibinfo {year} {2019})}\BibitemShut {NoStop}%
	\bibitem [{\citenamefont {Cocemasov}\ \emph {et~al.}(2013)\citenamefont
		{Cocemasov}, \citenamefont {Nika},\ and\ \citenamefont
		{Balandin}}]{Balandin}%
	\BibitemOpen
	\bibfield  {author} {\bibinfo {author} {\bibfnamefont {A.~I.}\ \bibnamefont
			{Cocemasov}}, \bibinfo {author} {\bibfnamefont {D.~L.}\ \bibnamefont {Nika}},
		\ and\ \bibinfo {author} {\bibfnamefont {A.~A.}\ \bibnamefont {Balandin}},\
	}\href@noop {} {\bibfield  {journal} {\bibinfo  {journal} {Phys. Rev. B}\
		}\textbf {\bibinfo {volume} {88}},\ \bibinfo {pages} {035428} (\bibinfo
		{year} {2013})}\BibitemShut {NoStop}%
	\bibitem [{\citenamefont {Ray}\ \emph {et~al.}(2016)\citenamefont {Ray},
		\citenamefont {Fleischmann}, \citenamefont {Weckbecker}, \citenamefont
		{Sharma}, \citenamefont {Pankratov},\ and\ \citenamefont
		{Shallcross}}]{RayTBG}%
	\BibitemOpen
	\bibfield  {author} {\bibinfo {author} {\bibfnamefont {N.}~\bibnamefont
			{Ray}}, \bibinfo {author} {\bibfnamefont {M.}~\bibnamefont {Fleischmann}},
		\bibinfo {author} {\bibfnamefont {D.}~\bibnamefont {Weckbecker}}, \bibinfo
		{author} {\bibfnamefont {S.}~\bibnamefont {Sharma}}, \bibinfo {author}
		{\bibfnamefont {O.}~\bibnamefont {Pankratov}}, \ and\ \bibinfo {author}
		{\bibfnamefont {S.}~\bibnamefont {Shallcross}},\ }\href@noop {} {\bibfield
		{journal} {\bibinfo  {journal} {Phys. Rev. B}\ }\textbf {\bibinfo {volume}
			{94}},\ \bibinfo {pages} {245403} (\bibinfo {year} {2016})}\BibitemShut
	{NoStop}%
	\bibitem [{\citenamefont {Chung}\ \emph
		{et~al.}(2018{\natexlab{b}})\citenamefont {Chung}, \citenamefont {Xu},\ and\
		\citenamefont {Chen}}]{Chung}%
	\BibitemOpen
	\bibfield  {author} {\bibinfo {author} {\bibfnamefont {T.-F.}\ \bibnamefont
			{Chung}}, \bibinfo {author} {\bibfnamefont {Y.}~\bibnamefont {Xu}}, \ and\
		\bibinfo {author} {\bibfnamefont {Y.~P.}\ \bibnamefont {Chen}},\ }\href
	{\doibase 10.1103/PhysRevB.98.035425} {\bibfield  {journal} {\bibinfo
			{journal} {Phys. Rev. B}\ }\textbf {\bibinfo {volume} {98}},\ \bibinfo
		{pages} {035425} (\bibinfo {year} {2018}{\natexlab{b}})}\BibitemShut
	{NoStop}%
	\bibitem [{\citenamefont {Polshyn}\ \emph {et~al.}(2019)\citenamefont
		{Polshyn}, \citenamefont {Yankowitz}, \citenamefont {Chen}, \citenamefont
		{Zhang}, \citenamefont {Watanabe}, \citenamefont {Taniguchi}, \citenamefont
		{Dean},\ and\ \citenamefont {Young}}]{Polshyn2019}%
	\BibitemOpen
	\bibfield  {author} {\bibinfo {author} {\bibfnamefont {H.}~\bibnamefont
			{Polshyn}}, \bibinfo {author} {\bibfnamefont {M.}~\bibnamefont {Yankowitz}},
		\bibinfo {author} {\bibfnamefont {S.}~\bibnamefont {Chen}}, \bibinfo {author}
		{\bibfnamefont {Y.}~\bibnamefont {Zhang}}, \bibinfo {author} {\bibfnamefont
			{K.}~\bibnamefont {Watanabe}}, \bibinfo {author} {\bibfnamefont
			{T.}~\bibnamefont {Taniguchi}}, \bibinfo {author} {\bibfnamefont {C.~R.}\
			\bibnamefont {Dean}}, \ and\ \bibinfo {author} {\bibfnamefont {A.~F.}\
			\bibnamefont {Young}},\ }\href {\doibase 10.1038/s41567-019-0596-3}
	{\bibfield  {journal} {\bibinfo  {journal} {Nature Physics}\ }\textbf
		{\bibinfo {volume} {15}},\ \bibinfo {pages} {1011} (\bibinfo {year}
		{2019})}\BibitemShut {NoStop}%
	\bibitem [{\citenamefont {Tomić}\ \emph {et~al.}(2021)\citenamefont {Tomić},
		\citenamefont {Rickhaus}, \citenamefont {Garcia-Ruiz}, \citenamefont {Zheng},
		\citenamefont {Portolés}, \citenamefont {Fal'ko}, \citenamefont {Watanabe},
		\citenamefont {Taniguchi}, \citenamefont {Ensslin}, \citenamefont {Ihn},\
		and\ \citenamefont {de~Vries}}]{Rickhaus_intervalley}%
	\BibitemOpen
	\bibfield  {author} {\bibinfo {author} {\bibfnamefont {P.}~\bibnamefont
			{Tomić}}, \bibinfo {author} {\bibfnamefont {P.}~\bibnamefont {Rickhaus}},
		\bibinfo {author} {\bibfnamefont {A.}~\bibnamefont {Garcia-Ruiz}}, \bibinfo
		{author} {\bibfnamefont {G.}~\bibnamefont {Zheng}}, \bibinfo {author}
		{\bibfnamefont {E.}~\bibnamefont {Portolés}}, \bibinfo {author}
		{\bibfnamefont {V.}~\bibnamefont {Fal'ko}}, \bibinfo {author} {\bibfnamefont
			{K.}~\bibnamefont {Watanabe}}, \bibinfo {author} {\bibfnamefont
			{T.}~\bibnamefont {Taniguchi}}, \bibinfo {author} {\bibfnamefont
			{K.}~\bibnamefont {Ensslin}}, \bibinfo {author} {\bibfnamefont
			{T.}~\bibnamefont {Ihn}}, \ and\ \bibinfo {author} {\bibfnamefont {F.~K.}\
			\bibnamefont {de~Vries}},\ }\href@noop {} {\enquote {\bibinfo {title}
			{Scattering between minivalleys in a moir\'e material},}\ } (\bibinfo {year}
	{2021}),\ \Eprint {http://arxiv.org/abs/2106.07805} {arXiv:2106.07805
		[cond-mat.mes-hall]} \BibitemShut {NoStop}%
	\bibitem [{\citenamefont {Krishna~Kumar}\ \emph {et~al.}(2017)\citenamefont
		{Krishna~Kumar}, \citenamefont {Chen}, \citenamefont {Auton}, \citenamefont
		{Mishchenko}, \citenamefont {Bandurin}, \citenamefont {Morozov},
		\citenamefont {Cao}, \citenamefont {Khestanova}, \citenamefont {Ben~Shalom},
		\citenamefont {Kretinin}, \citenamefont {Novoselov}, \citenamefont {Eaves},
		\citenamefont {Grigorieva}, \citenamefont {Ponomarenko}, \citenamefont
		{Fal{\textquoteright}ko},\ and\ \citenamefont {Geim}}]{Rosh_BZ}%
	\BibitemOpen
	\bibfield  {author} {\bibinfo {author} {\bibfnamefont {R.}~\bibnamefont
			{Krishna~Kumar}}, \bibinfo {author} {\bibfnamefont {X.}~\bibnamefont {Chen}},
		\bibinfo {author} {\bibfnamefont {G.~H.}\ \bibnamefont {Auton}}, \bibinfo
		{author} {\bibfnamefont {A.}~\bibnamefont {Mishchenko}}, \bibinfo {author}
		{\bibfnamefont {D.~A.}\ \bibnamefont {Bandurin}}, \bibinfo {author}
		{\bibfnamefont {S.~V.}\ \bibnamefont {Morozov}}, \bibinfo {author}
		{\bibfnamefont {Y.}~\bibnamefont {Cao}}, \bibinfo {author} {\bibfnamefont
			{E.}~\bibnamefont {Khestanova}}, \bibinfo {author} {\bibfnamefont
			{M.}~\bibnamefont {Ben~Shalom}}, \bibinfo {author} {\bibfnamefont {A.~V.}\
			\bibnamefont {Kretinin}}, \bibinfo {author} {\bibfnamefont {K.~S.}\
			\bibnamefont {Novoselov}}, \bibinfo {author} {\bibfnamefont {L.}~\bibnamefont
			{Eaves}}, \bibinfo {author} {\bibfnamefont {I.~V.}\ \bibnamefont
			{Grigorieva}}, \bibinfo {author} {\bibfnamefont {L.~A.}\ \bibnamefont
			{Ponomarenko}}, \bibinfo {author} {\bibfnamefont {V.~I.}\ \bibnamefont
			{Fal{\textquoteright}ko}}, \ and\ \bibinfo {author} {\bibfnamefont {A.~K.}\
			\bibnamefont {Geim}},\ }\href {\doibase 10.1126/science.aal3357} {\bibfield
		{journal} {\bibinfo  {journal} {Science}\ }\textbf {\bibinfo {volume}
			{357}},\ \bibinfo {pages} {181} (\bibinfo {year} {2017})}\BibitemShut
	{NoStop}%
	\bibitem [{\citenamefont {Wallbank}\ \emph {et~al.}(2019)\citenamefont
		{Wallbank}, \citenamefont {{Krishna Kumar}}, \citenamefont {Holwill},
		\citenamefont {Wang}, \citenamefont {Auton}, \citenamefont {Birkbeck},
		\citenamefont {Mishchenko}, \citenamefont {Ponomarenko}, \citenamefont
		{Watanabe}, \citenamefont {Taniguchi}, \citenamefont {Novoselov},
		\citenamefont {Aleiner}, \citenamefont {Geim},\ and\ \citenamefont
		{Fal'ko}}]{Wallbank2019}%
	\BibitemOpen
	\bibfield  {author} {\bibinfo {author} {\bibfnamefont {J.~R.}\ \bibnamefont
			{Wallbank}}, \bibinfo {author} {\bibfnamefont {R.}~\bibnamefont {{Krishna
					Kumar}}}, \bibinfo {author} {\bibfnamefont {M.}~\bibnamefont {Holwill}},
		\bibinfo {author} {\bibfnamefont {Z.}~\bibnamefont {Wang}}, \bibinfo {author}
		{\bibfnamefont {G.~H.}\ \bibnamefont {Auton}}, \bibinfo {author}
		{\bibfnamefont {J.}~\bibnamefont {Birkbeck}}, \bibinfo {author}
		{\bibfnamefont {A.}~\bibnamefont {Mishchenko}}, \bibinfo {author}
		{\bibfnamefont {L.~A.}\ \bibnamefont {Ponomarenko}}, \bibinfo {author}
		{\bibfnamefont {K.}~\bibnamefont {Watanabe}}, \bibinfo {author}
		{\bibfnamefont {T.}~\bibnamefont {Taniguchi}}, \bibinfo {author}
		{\bibfnamefont {K.~S.}\ \bibnamefont {Novoselov}}, \bibinfo {author}
		{\bibfnamefont {I.~L.}\ \bibnamefont {Aleiner}}, \bibinfo {author}
		{\bibfnamefont {A.~K.}\ \bibnamefont {Geim}}, \ and\ \bibinfo {author}
		{\bibfnamefont {V.~I.}\ \bibnamefont {Fal'ko}},\ }\href
	{https://doi.org/10.1038/s41567-018-0278-6} {\bibfield  {journal} {\bibinfo
			{journal} {Nature Physics}\ }\textbf {\bibinfo {volume} {15}},\ \bibinfo
		{pages} {32} (\bibinfo {year} {2019})}\BibitemShut {NoStop}%
\end{thebibliography}

\end{document}